\documentclass[useAMS,usenatbib,usegraphicx]{mn2e}
\usepackage{graphicx,amssymb}
\usepackage{color}

\title[The Blazhko modulation of XY~And and UZ~Vir]
      {The light-curve modulation of XY~And and UZ~Vir --\\ Two Blazhko RR Lyrae stars with additional frequencies\thanks{Based on observations collected with the automatic 60-cm telescope of Konkoly Observatory, Budapest, Sv\'abhegy}}

\author[\'A. S\'odor et al.]
       {\'A. S\'odor$^{1}$\thanks{E-mail: sodor@konkoly.hu}, G. Hajdu$^{1}$, J. Jurcsik$^{1}$, B. Szeidl$^{1}$, K. Posztob\'anyi$^{2}$, Zs. Hurta$^{2}$, \and B. Belucz$^3$, E. Kun$^4$
        \\
        $^{1}$Konkoly Observatory, Research Centre for Astronomy and Earth Sciences, \\
        \ \ Hungarian Academy of Sciences, Budapest, Hungary\\
        $^{2}$Visiting astronomer at Konkoly Observatory\\
        $^{3}$E\"otv\"os University, Dept. of Astronomy, H-1518 Budapest PO Box 49, Hungary\\                                                                                               
        $^{4}$Department of Experimental Physics and Astronomical Observatory, University of Szeged, 6720 Szeged, D\'om t\'er 9, Hungary\\
       }

\begin{document}

\date{Accepted 2010 ..... Received 2010 ...}

\pagerange{\pageref{firstpage}--\pageref{lastpage}} \pubyear{2012}

\maketitle

\label{firstpage}

\begin{abstract}
A thorough analysis of multicolour CCD observations of two modulated RRab-type variables, XY~And and UZ~Vir is presented. These Blazhko stars show relatively simple light-curve modulation with the usual multiplet structures in their Fourier spectra. One additional, independent frequency with linear-combination terms of the pulsation frequency is also detected in the residual spectrum of each of the two stars. The amplitude and phase relations of the triplet components are studied in detail. Most of the epoch-independent phase differences show a slight, systematic colour dependence, however, these trends have the opposite sign in the two stars. The mean values of the global physical parameters and their changes with Blazhko phase are determined utilizing the Inverse Photometric Method (IPM). The modulation properties and the IPM results are compared for the two variables. The pulsation period of XY~And is the shortest when its pulsation amplitude is the highest, while UZ~Vir has the longest pulsation period at this phase of the modulation. Despite this opposite behaviour, the phase relations of their mean-physical-parameter variations are similar. These results are not in accord with the predictions of the Blazhko model of Stothers (2006, ApJ, 652, 643).
\end{abstract}


\section{Introduction}

In spite of the recently increased research activity in the field, we still do not have a satisfactory model for the phenomenon of the light-curve modulation of RR~Lyrae stars, the so-called Blazhko effect, that would be consistent with the wide variety of the ever increasing collection of observations. To find such a model, not only the phenomenology of the light-curve variation have to be known; the modulational changes in the atmospheric physical parameters of these stars also impose constraints on the potential Blazhko models. The brightness variations of Blazhko RR~Lyrae stars can be studied with unprecedented detail and precision through the data of {\it CoRoT} and {\it Kepler} space telescopes in a single photometric band, and these studies led to new explanations of the phenomenon with the detection of half-integer frequencies \citep{bk12,kmsz}.
However, to derive physical parameters, standard multicolour or spectroscopic observations are needed. Since the modulation periods of Blazhko stars span a wide range from about one week to many hundred days, detailed spectroscopic studies with sufficiently large-aperture telescopes are rather difficult to attain. Therefore, moderate-size earth-based multicolour photometric telescopes still have an important role in studying the Blazhko effect.

In the context of the Konkoly Blazhko Survey I and II \citep{kbs1,aspc}, we have obtained multicolour light curves of bright, northern, fundamental-mode Blazhko RR~Lyrae stars. These data have already been successfully utilized to derive the variations in the atmospheric parameters of six modulated stars (see sect. 6 in \citealt{rzl} and references therein).

According to the explanation of \cite{stothers}, the Blazhko effect is `a direct consequence of a gradual strengthening and weakening of turbulent convection in the stellar envelope'. Though this model has been seriously criticized on theoretical grounds by \cite{ko09,sm} and \cite{mol}, it is also important to confront the predictions of the Stothers-model with observational results.
   The model predicts that the phase relation between the amplitude and period changes depends on the mean physical properties of the star and it is also in direct connection with the phase relations of the physical parameter variations during the Blazhko cycle. The present paper aims to test if there is any connection between the phase relation of the amplitude and period (phase) variations and the mean physical parameters and their variations during the Blazhko cycle as proposed by \cite{stothers,stothers2011}. Therefore, we have selected two stars to study their Blazhko modulation in detail, XY~And ($\alpha_{2000} = 01^{\rm h}26^{\rm m}42.\!\!^{\rm s}43$, $\delta_{2000} = +34{\degr}04'06.\!\!^{\prime\prime}9$, $P=0.3987$~d, $P_{\mathrm mod}=41.4$~d) and UZ~Vir ($\alpha_{2000} = 13^{\rm h}08^{\rm m}44.\!\!^{\rm s}32$, $\delta_{2000} = +13{\degr}24'08.\!\!^{\prime\prime}4$, $P=0.4593$~d, $P_{\mathrm mod}=68.2$~d), which show different phase relations between their amplitude and period (phase) variations. XY~And and UZ~Vir were extensively observed in the course of the Konkoly Blazhko Survey I \citep{kbs1}, but the data and the results of the light-curve analysis have not been published yet.

\section{Observations}

\begin{table}
  \centering
  \caption{Log of the CCD observations of XY~And and UZ~Vir obtained with the 60-cm automatic telescope of Konkoly Observatory.\label{tbl:obslog}}
  \begin{tabular}{ccclcc}
    \hline
    Object & From & To & Filter & Nights & Data \\
           & \multicolumn{2}{c}{(JD\,$-$\,2\,450\,000)} & & & points \\
    \hline
    XY And & 4390 & 4857 & \ \ $V$            & 64 & 2611 \\
           &      &      & \ \ $I_\mathrm{C}$ & 64 & 2610 \\
    UZ Vir & 4504 & 4978 & \ \ $B$            & 72 & 1722 \\
           &      &      & \ \ $V$            & 70 & 1695 \\
           &      &      & \ \ $I_\mathrm{C}$ & 70 & 1676 \\
    \hline
  \end{tabular}
\end{table}

\begin{table}
\centering
\caption{Identification of the variables and their comparison stars.\label{tbl:comp}}
  \begin{tabular}{ccc}
  \hline
Object & 2MASS ID         & Comp. star 2MASS ID\\
\hline
XY~And & 01264243+3404068 & 01270016+3404307\\
UZ~Vir & 13084432+1324084 & 13082756+1322403\\
\hline
\end{tabular}
\end{table}

\begin{figure*}
\centering
\includegraphics[width=7.8 cm]{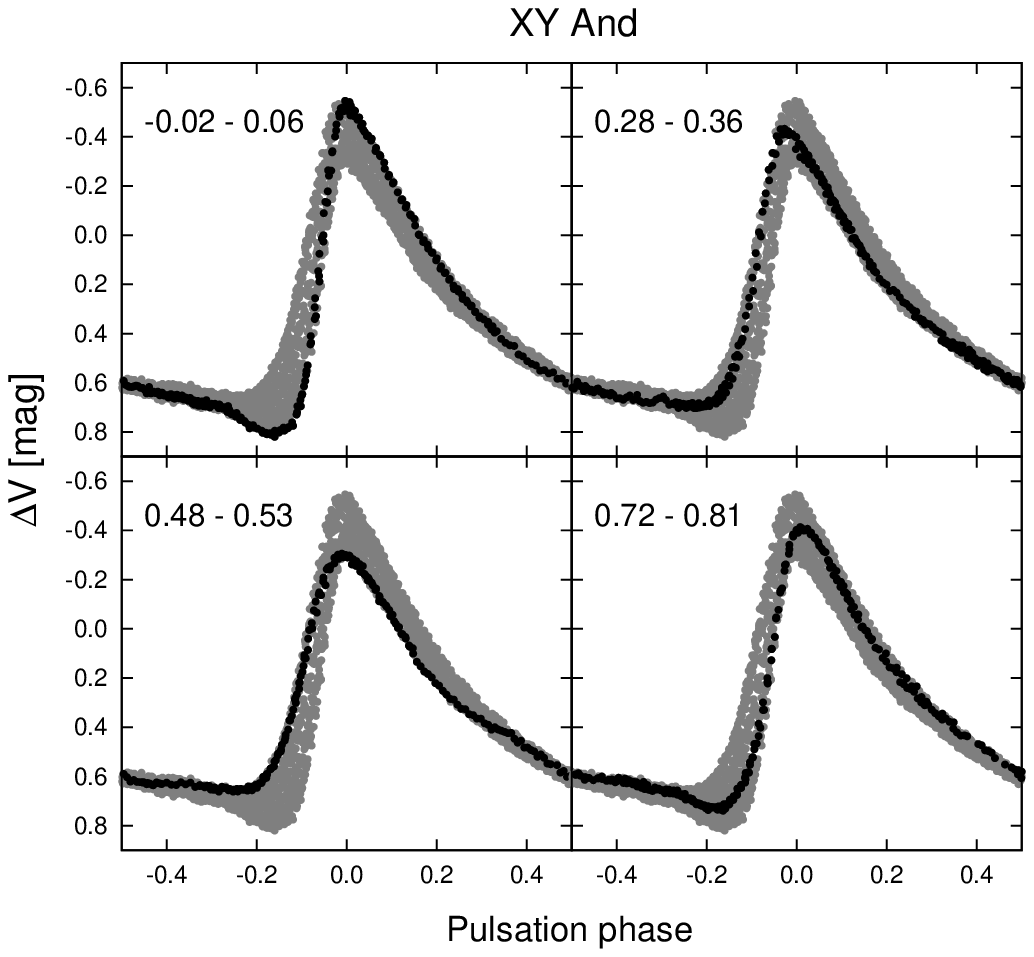}
\includegraphics[width=7.8 cm]{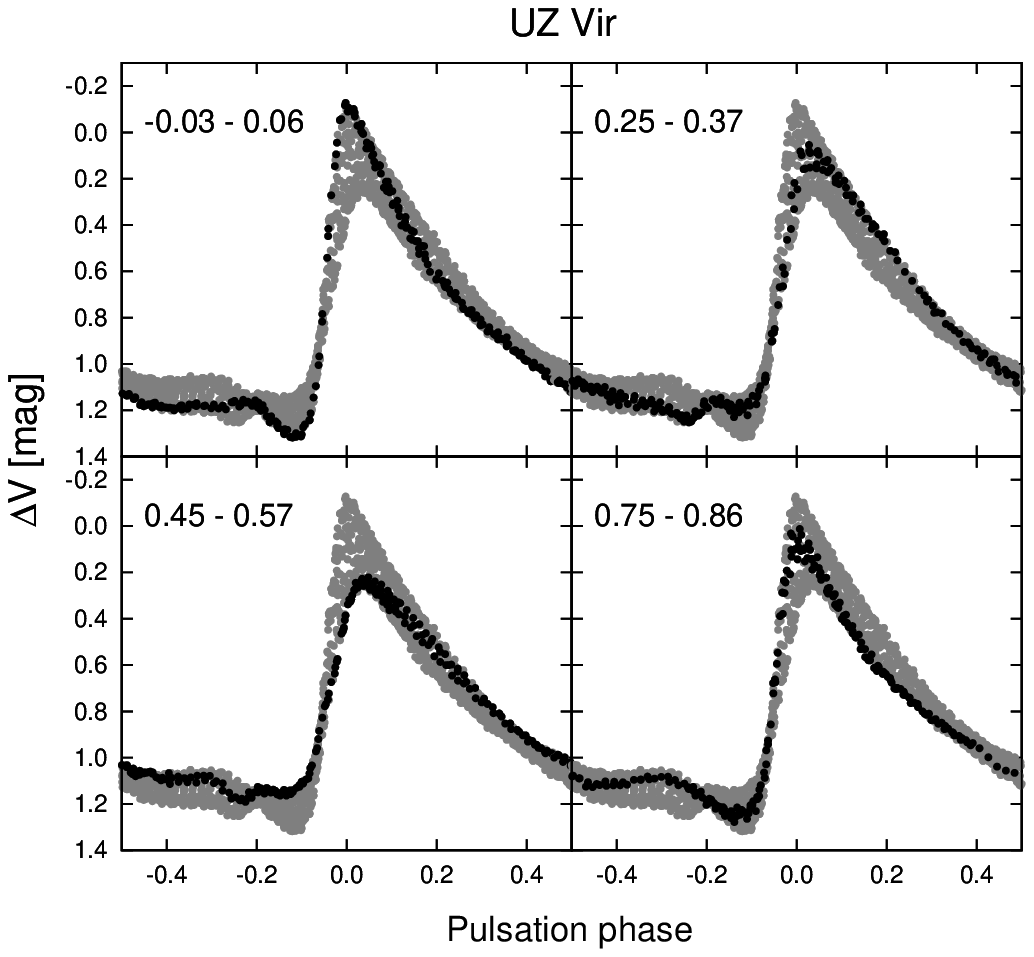}
\caption{Folded $V$ light curves of XY~And and UZ~Vir in selected Blazhko phase bins. The gray dots indicate all the measurements. In the top-left corners, the Blazhko-phase ranges of the highlighted data are given, according to the Blazhko ephemerides, Eqs.~\ref{eq:blephem_xy} and \ref{eq:blephem_uz}. The rising branch of XY~And shows a significant phase variation ($\sim$0.1 pulsation phase) while the fix point on the rising branch of UZ~Vir indicates that amplitude modulation dominates its light-curve variation. The phase of the maximum brightness varies, however, within about the same, $\pm 0.02$ pulsation-phase range in both stars, as shown in Fig.~\ref{fig:tojas}.}
\label{fig:lc}
\end{figure*}

Multicolour CCD observations of XY~And and UZ~Vir were obtained between February 2008 and May 2009, and between October 2007 and January 2009, respectively, with the 60-cm automatic telescope of Konkoly Observatory at Sv\'abhegy, Budapest, equipped with a $750 \times 1100$ Wright Instruments CCD camera. In the case of UZ~Vir, Johnson--Cousins $BVI_{\mathrm{C}}$ filters were used, while for the $\sim0.7$\,mag fainter XY~And, the observations were made only in the $V$ and $I_{\mathrm{C}}$ bands in order to assure a reasonably short duty cycle. In each band, the total numbers of the observations are about 2600 and 1700 for XY~And and UZ~Vir, respectively (see Table~\ref{tbl:obslog} for the log of observations.) Standard CCD calibration and aperture photometry were performed on all the object frames using IRAF\footnote{{\sc IRAF} is distributed by the National Optical Astronomy Observatories, which are operated by the Association of Universities for Research in Astronomy, Inc., under cooperative agreement with the National Science Foundation.} packages. The relative magnitudes of the variables were measured against the nearby comparison stars identified in Table~\ref{tbl:comp}. The relative magnitudes were transformed to the Johnson--Cousins system utilizing standard procedures.

Fig.~\ref{fig:lc} shows the light curves of the variables folded with the pulsation period at selected Blazhko phases.

The light-curve data of XY~And and UZ~Vir are available as supporting information with the online version of this paper. Tables~\ref{tbl:dataxyav}--\ref{tbl:datauzvi} in Appendix~\ref{appendix:electronic} give information regarding the form and content of the data.

\begin{figure*}
\centering
\includegraphics[width=7.2cm]{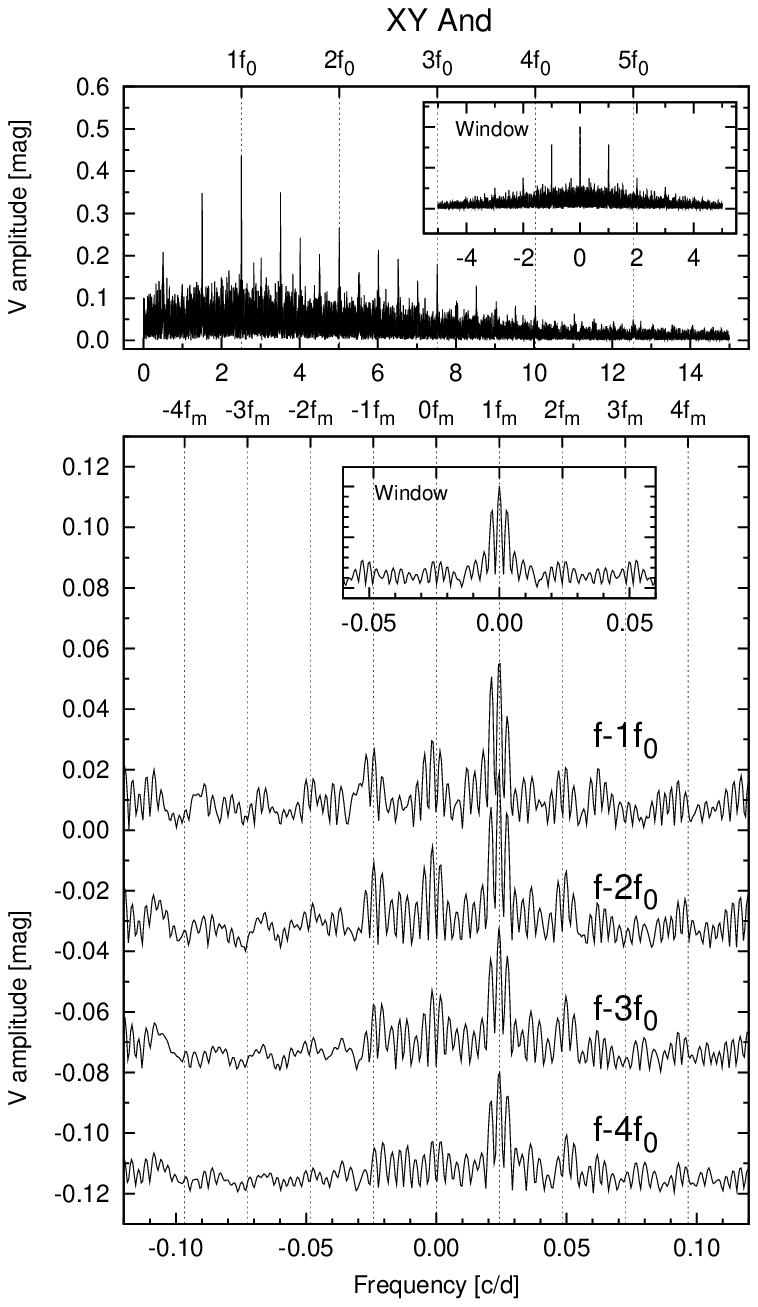}
\includegraphics[width=7.2cm]{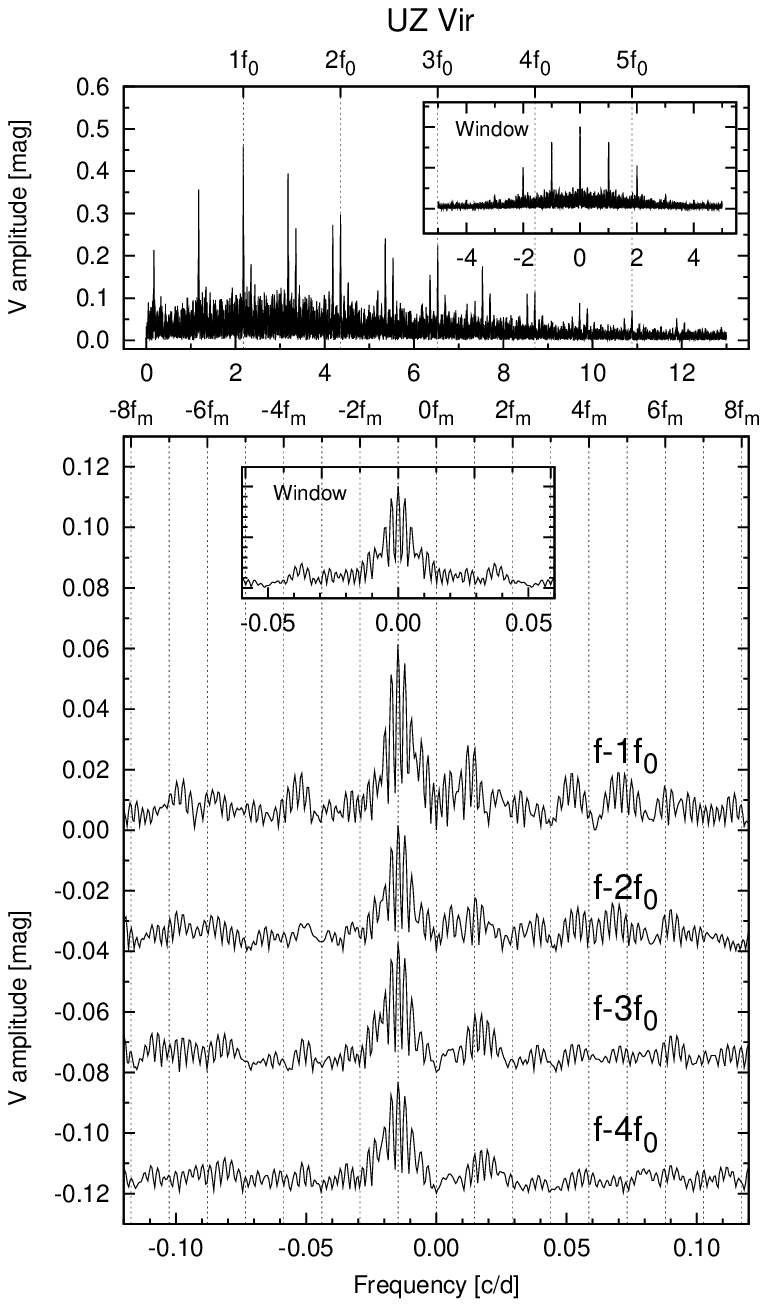}
\caption{Fourier amplitude spectra of the $V$ light curves of XY~And and UZ~Vir (top panels). The residual spectra are plotted in the bottom panels around the first four harmonics of the pulsation frequency after the removal of the detected pulsation components ($kf_0$, $k=\left\{1, ..., 13\right\}$ for XY~And and $k=\left\{1, ..., 25\right\}$ for UZ~Vir). The spectrum segments are artificially shifted for clarity. Peaks belonging to the modulation ($kf_0\pm f_\mathrm{m}$) are clearly identified. The window functions, shown in the inserts, demonstrate that no high-amplitude alias frequencies are present besides the daily and yearly aliases.
\label{fig:sp1}
}
\end{figure*}

\section{Frequency analysis}
\label{sect:freq}

The mathematical description of the light curves is sought by Discrete Fourier Transform \citep{de75} as implemented in the {\sc mufran} package \citep{mufran}, the \hbox{\sc gnuplot} utilities\footnote{http://www.gnuplot.info/} and a non-linear fitting algorithm \citep{nlfit}.

A peak in the Fourier spectrum is accepted as an independent frequency of the star and is included into the solution if its amplitude exceeds 3.5$\sigma$ in at least two photometric bands ($\sigma$ is defined as the local mean level of the residual spectrum). For linear combination terms of the independent frequencies, the selection criterion is that its amplitude exceeds 2.5$\sigma$ in at least two bands or 3$\sigma$ in one band.

\subsection{Pulsation and modulation components}

The top panels of Fig.~\ref{fig:sp1} show the spectra of the $V$ light curves of XY~And and UZ~Vir. The bottom panels display the residual spectra after prewhitening for the pulsation frequency and its harmonics ($kf_0$). The window functions are plotted in the inserts. Side-peaks belonging to the modulation triplet frequencies  ($kf_0\pm f_\mathrm{m}$) with significantly different amplitudes of the two components are clearly identified in the residual spectrum of the variables. After prewhitening for the pulsation frequency, its harmonics, the modulation frequency and their identified linear-combination terms, no further triplet (multiplet) structure, indicative of a secondary modulation, is found around the pulsation frequency and its harmonics in any of the stars.

\begin{figure}
\centering
\includegraphics[width=7.1 cm]{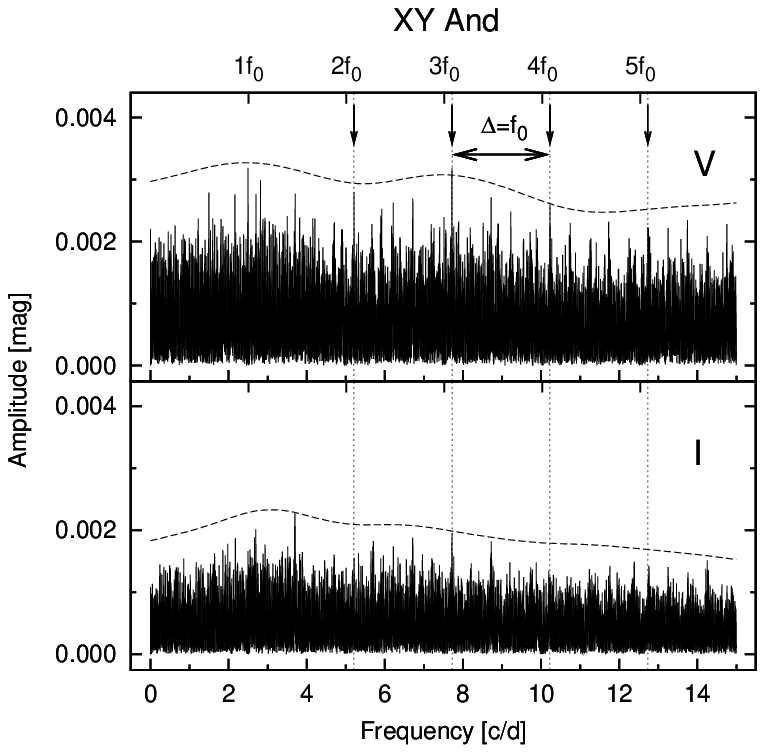}
\vskip4mm
\includegraphics[width=7.1 cm]{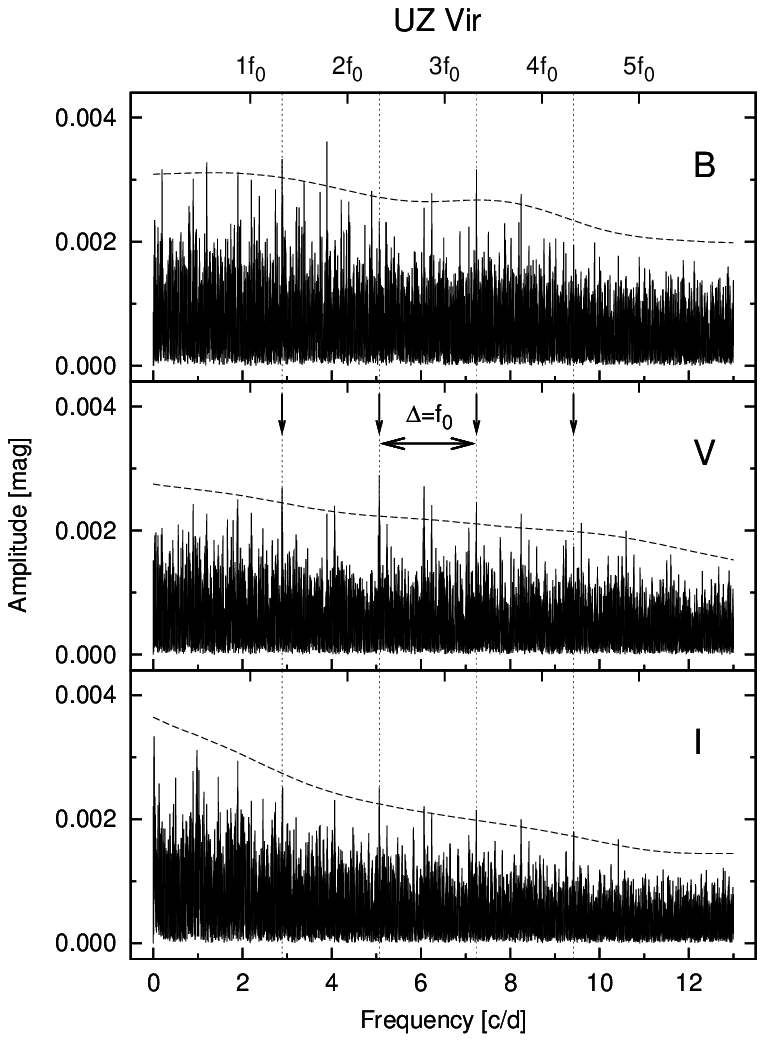}
\caption{Residual amplitude spectra of XY~And and UZ~Vir in the observed bands after the removal of the pulsation frequency, its harmonics, the modulation frequency and their linear combinations. Dashed lines show 3.5 times the mean level of the spectra (3.5$\sigma$). A regular pattern is evident in the spectra of both stars: peaks are present at 5.21, 7.72, 10.23 and 12.74\,cd$^{-1}$, and at 2.89, 5.06, 7.24 and 9.42\,cd$^{-1}$ in the case of XY~And and UZ~Vir, respectively. These frequency series (indicated by vertical dotted lines) are separated by the pulsation frequency. 
\label{fig:sp2}}
\end{figure}

\subsection{Detection of additional frequencies}
\label{subsect:addfr}

\begin{table}
  \centering
  \caption{The sequences of identified additional frequencies in XY~And and UZ~Vir. The ratios with the pulsation frequency ($f_{0}/f_{i}'$), the amplitudes and the significance are given. For UZ~Vir, the $\pm$1 cd$^{-1}$ alias frequencies are listed as well.}
  \label{tbl:addfr}
  \begin{tabular}{@{\extracolsep{-3pt}}lrllclclc}
    \hline
           &\multicolumn{8}{c}{XY~And}       \\
           &freq.       & $f_{0}/f_{i}'$ & $A(B)$                  & $\sigma$                &$A(V)$  & $\sigma$  & $A(I)$ & $\sigma$ \\
           &(cd$^{-1}$) &                & (mag)                   &                         &(mag)   &           & (mag) \\
    \hline
$f_{1}'$   & 5.2122     & 0.481          & \multicolumn{1}{c}{---} & \multicolumn{1}{c}{---} & 0.0028 & 3.3       & 0.0013 & 2.2 \\
$f_{2}'$   & 7.7202     & 0.325          & \multicolumn{1}{c}{---} & \multicolumn{1}{c}{---} & 0.0032 & 3.6       & 0.0019 & 3.5 \\
$f_{3}'$   & 10.2282    & 0.245          & \multicolumn{1}{c}{---} & \multicolumn{1}{c}{---} & 0.0026 & 3.5       & 0.0011 & 2.2 \\
$f_{4}'$   & 12.7362    & 0.197          & \multicolumn{1}{c}{---} & \multicolumn{1}{c}{---} & 0.0022 & 3.1       & 0.0011 & 2.3 \\
\hline
\hline
           &\multicolumn{8}{c}{UZ~Vir}       \\
\hline
$f_{1}'$   &\ \,2.8880  & 0.754          & 0.0033                  & 3.8                     & 0.0027 & 3.9       & 0.0019 & 2.4 \\
$f_{2}'$   &\ \,5.0648  & 0.430          & 0.0023                  & 3.0                     & 0.0029 & 4.5       & 0.0025 & 3.9 \\
$f_{3}'$   &\ \,7.2416  & 0.301          & 0.0032                  & 4.2                     & 0.0025 & 4.2       & 0.0021 & 3.7 \\
$f_{4}'$   &\ \,9.4184  & 0.231          & 0.0019                  & 2.8                     & 0.0017 & 3.0       & 0.0018 & 3.7 \\
\hline
$f_{1}'^-$ &\ \,1.8880  & 1.153          & 0.0032                  & 3.9                     & 0.0022 & 3.3       & 0.0015 & 2.1 \\
$f_{2}'^-$ &\ \,4.0648  & 0.536          & 0.0021                  & 2.8                     & 0.0025 & 4.0       & 0.0022 & 3.6 \\
$f_{3}'^-$ &\ \,6.2416  & 0.349          & 0.0027                  & 3.6                     & 0.0021 & 3.6       & 0.0020 & 3.7 \\
$f_{4}'^-$ &\ \,8.4184  & 0.259          & 0.0018                  & 3.0                     & 0.0016 & 2.9       & 0.0017 & 3.8 \\
\hline
$f_{1}'^+$ &\ \,3.8880  & 0.560          & 0.0022                  & 2.5                     & 0.0021 & 2.9       & 0.0018 & 2.0 \\
$f_{2}'^+$ &\ \,6.0648  & 0.359          & 0.0018                  & 2.2                     & 0.0022 & 3.3       & 0.0023 & 3.3 \\
$f_{3}'^+$ &\ \,8.2416  & 0.264          & 0.0023                  & 3.0                     & 0.0019 & 3.1       & 0.0016 & 2.7 \\
$f_{4}'^+$ &   10.4184  & 0.209          & 0.0017                  & 2.3                     & 0.0017 & 2.9       & 0.0013 & 2.5 \\
    \hline
  \end{tabular}
\end{table}

Fig.~\ref{fig:sp2} shows the residual spectra of XY~And and UZ~Vir in the observed bands after the removal of the detected multiplets. Both spectra have a regular pattern of peaks with separations of the pulsation frequency, $f_0$. The frequency values, the $f_0/f$ frequency ratios, the amplitudes and their significance are given in Table~\ref{tbl:addfr}. Most probably, one of these frequencies is an additional (radial or non-radial) mode, while the other peaks are linear combinations of this mode and the pulsation frequency. For UZ Vir, $\pm 1$\,cd$^{-1}$ alias solutions for the identification of the `base' frequency of the additional frequency series cannot be excluded either.

\subsubsection{The additional frequency of XY~And}

We accept the second of the series listed in Table~\ref{tbl:addfr} ($f_{2}'=7.7202$) as an independent, additional frequency, because it has the highest amplitude both  in the $V$ and $I_\mathrm{C}$ bands. This frequency may be identified as a non-radial mode, as model calculations \citep{bk01,kmsz} indicate that neither normal nor `strange'-modes have positive growth-rates with frequency ratio around 0.325, which is between the regime of the fifth and the sixth radial overtones \citep[see figure 3. in][]{kmsz}.

The explanation that these peaks belong to a secondary modulation series with $\sim 0.196$\,cd$^{-1}$ separation from the harmonics of the pulsation cannot be excluded either. However, both the corresponding extremely short modulation period ($\sim5.01$\,d) and the lack of the triplet components on the other side of the pulsation frequency make this explanation less likely.

\subsubsection{The additional frequency of UZ~Vir}

In the case of UZ~Vir, any of the $f_1'$, $f_2'$ and $f_3'$ frequencies can be considered as the base-frequency of the additional frequency series; $f_1'$ and $f_3'$ are strong in the $B$ band, while the amplitude of $f_2'$ is the highest in the $V$ and $I_\mathrm{C}$ bands. When fitting the final frequency solution to the light curves, $f_2'$ was treated as a mathematically independent frequency, but this technical choice does not affect our subsequent results and conclusions.

As Fig.~\ref{fig:sp2} shows, the $\pm 1$\,cd$^{-1}$ alias frequencies of the additional-frequency series of UZ~Vir have similar amplitudes as the selected ones, which poses further ambiguities. Fitting the light curve of UZ~Vir with all the detected frequencies and with the series of  $f_i'$ listed in the first block of Table~\ref{tbl:addfr}, the rms values of the residuals are 0.0112, 0.0093 and 0.0096\,mag in the $B$, $V$ and $I_{\mathrm{C}}$ bands, respectively. The solutions including the $f_{i}'^+$ and $f_{i}'^-$ alias frequency series yield marginally higher, 0.0113, 0.0095, 0.0097 and 0.0114, 0.0094, 0.0097\,mag rms of the residuals. Moreover, all the amplitudes of the $f_1', f_2'$, $f_3'$ and $f_4'$ terms in the $B$, $V$ and $I_{\mathrm{C}}$ bands are higher than the corresponding amplitudes of the possible alias solutions. These results support our choice of the true additional-frequency series of UZ~Vir.

The frequency at 2.888\,cd$^{-1}$ ($f_1'$) has a period ratio of 0.754 with $f_0$.  Although this is significantly larger than the observed $f_0/f_1$ period ratios of double-mode RR Lyrae stars, an open question is, whether in special, resonant cases, a low-amplitude, first-overtone mode might really be excited at such a large frequency ratio (Z. Koll\'ath, private comm.). The frequency of the other possible base-component of this additional frequency series ($f_2'=5.065$ or $f_3'=7.242$) might be identified as the third/fourth and the sixth-order radial mode or a non-radial mode. However, as RR Lyr models show that radial modes between the third and eights radial orders are strongly damped, therefore this explanation seems to be unlikely. It is important to note here also that, according to the  study of \cite{dc}, non-radial modes have the highest chances for excitation at close proximity of the radial modes. The explanation of the additional-frequency series as a secondary modulation is rejected in this case, because the peaks are at a distance of 0.711\,cd$^{-1}$ from the pulsation harmonics, which corresponds to an unlikely short modulation period of only $\sim1.41$\,d.

\subsection{The final frequency solution}

None of the residual spectra shows any further significant peak after the removal of the frequency components identified in the previous subsections. No frequency component with amplitude larger than $3.5\sigma$ in at least one band is detected within the 0.05 d$^{-1}$ vicinity of the positions of the half integer pulsation frequencies ($kf_0/2, k=1,3,5,7,9$) in any of the two stars.

The final frequency solution of XY~And includes 57 components: 13 harmonics of the pulsation frequency ($kf_0$), the modulation frequency ($f_\mathrm{m}$), 24 linear-combination frequencies belonging to the triplets ($kf_0\pm f_\mathrm{m}$), 15 quintuplet components ($kf_0\pm 2f_\mathrm{m}$), the additional frequency ($f_{2}'$ in Table~\ref{tbl:addfr}) and 3 linear combinations of the additional frequency and the pulsation frequency. The rms values of the residuals are 0.011 and 0.008\,mag in the $V$ and $I_\mathrm{C}$ bands, respectively.

As for UZ~Vir, the total number of the identified frequency components is 84, of which 25 are harmonics of the pulsation frequency, one is the modulation frequency, 42 are triplet components, 11 are quintuplet components, one is a septuplet peak ($6f_0 - 3f_\mathrm{m}$),  another one is the additional frequency, and three linear combinations of the additional frequency and the pulsation frequency are also included. The rms values of the residuals are 0.011, 0.009 and 0.010\,mag in the $B$, $V$ and $I_\mathrm{C}$ bands, respectively.

The accepted values of the independent frequencies ($f_0$, $f_\mathrm{m}$, $f_2'$) are determined by a non-linear Fourier fit to the $V$ light curves of the stars, including all the identified linear combination components with locked frequencies. These frequency values, the corresponding periods and their errors are listed in Table~\ref{tbl:freq}. The tables containing the full light-curve solutions (the frequency components, amplitudes and phases in each band) are available as supporting information with the electronic version of this article. Tables~\ref{tbl:freqxya} and \ref{tbl:frequzv} are samples of these electronic data and give information regarding their form and content.

\begin{table}
  \centering
  \caption{Pulsation, modulation and additional frequencies, the corresponding periods and their errors (in parentheses, in the unit of the last digit) of XY~And and UZ~Vir.\label{tbl:freq}}
  \begin{tabular}{lll}
    \hline
                                    & \multicolumn{1}{c}{Frequency (cd$^{-1}$)}& \multicolumn{1}{c}{Period (d)}\\
                                    &\multicolumn{2}{c}{XY And}  \\
    \hline
    pulsation ($f_0$, $P_0$)        & 2.5079961(6) &\ \,0.3987247(1) \\
    modulation ($f_\mathrm{m}$, $P_\mathrm{m}$)   & 0.02417(3)   &41.37(5)     \\
    additional ($f_2'$, $P_2'$)     & 7.7202(6)    &\ \,0.12953(1)   \\
    \hline
                                    &\multicolumn{2}{c}{UZ Vir} \\
    \hline
    pulsation ($f_0$, $P_0$)        & 2.1767879(4) &\ \,0.4593925(1)\\
    modulation ($f_\mathrm{m}$, $P_\mathrm{m}$)   & 0.014654(3)  &68.24(1)\\
    additional ($f_2'$, $P_2'$)     & 5.06481(6)   &\ \,0.197441(2)\\
    \hline
  \end{tabular}
\end{table}

\begin{table*}
\caption{Frequencies, amplitudes and phases of the $V$ and $I_\mathrm{C}$ light-curve solutions (epoch $=54390.0$, sine decomposition) of XY~And. The full table is available as supporting information with the online version of this paper.
 \label{tbl:freqxya}
}
  \begin{tabular}{lcllll}
\hline
\multicolumn{6}{c}{XY~And}\\
\hline
                     &               &\multicolumn{2}{c}{$V$} &\multicolumn{2}{c}{$I_\mathrm{C}$}      \\
identification       &$f$            &amplitude    &phase     &amplitude    &phase\\
                     & (cd$^{-1}$)   & (mag)       & (rad)    &(mag)        &(rad)\\
\hline
$f_0$                &2.507996       &0.4263(3)    &5.275(1)  &0.2574(3)    &5.139(1)\\
$2f_0$               &5.015992       &0.2044(3)    &0.275(2)  &0.1267(3)    &0.241(2)\\
...\\
$f_\mathrm{m}$       &0.024171       &0.0101(3)    &3.11(3)   &0.0062(3)    &3.34(4)\\
$f_0+f_\mathrm{m}$   &2.532167       &0.0528(3)    &6.107(6)  &0.0320(3)    &6.127(8)\\
$2f_0+f_\mathrm{m}$  &5.040163       &0.0506(3)    &1.190(7)  &0.0317(3)    &1.278(8)\\
...\\
$f_0-f_\mathrm{m}$   &2.483825       &0.0225(3)    &2.07(2)   &0.0138(3)    &2.19(2)\\
$2f_0-f_\mathrm{m}$  &4.991821       &0.0239(3)    &3.47(1)   &0.0150(3)    &3.53(2)\\
...\\
$3f_0+2f_\mathrm{m}$ &7.572330       &0.0018(3)    &2.9(2)    &0.0016(3)    &3.0()\\
$4f_0+2f_\mathrm{m}$ &10.08033       &0.0016(3)    &4.9(2)    &0.0009(3)    &5.6(3)\\
...\\
$f_0-2f_\mathrm{m}$  &2.459654       &0.0030(3)    &5.5(1)    &0.0019(3)    &5.8(1)\\
$2f_0-2f_\mathrm{m}$ &4.967650       &0.0015(4)    &5.9(2)    &0.0010(3)    &5.2(3)\\
...\\
$f_1'=f_2'-f_0$      &5.212223       &0.0026(4)    &5.5(1)    &0.0012(3)    &0.1(2)\\
$f_2'$               &7.720219       &0.0029(4)    &0.4(1)    &0.0019(3)    &0.8(2)\\
$f_3'=f_2'+f_0$      &10.22822       &0.0021(4)    &2.7(2)    &0.0006(3)    &2.2(5)\\
$f_4'=f_2'+2f_0$     &12.73621       &0.0017(4)    &4.0(2)    &0.0009(3)    &4.2(3)\\
\hline
\end{tabular}
\end{table*}

\begin{table*}
\caption{Frequencies, amplitudes and phases of the $B$, $V$ and $I_\mathrm{C}$ light-curve solutions (epoch $=54404.0$, sine decomposition) of UZ~Vir. The full table is available as supporting information with the online version of this paper.
 \label{tbl:frequzv}
}
  \begin{tabular}{lcllllll}
\hline
\multicolumn{8}{c}{UZ~Vir}\\
\hline
                      &               &\multicolumn{2}{c}{$B$} &\multicolumn{2}{c}{$V$} &\multicolumn{2}{c}{$I_\mathrm{C}$}      \\
identification        &$f$            &amplitude     &phase    &amplitude    &phase     &amplitude    &phase\\
                      &  (cd$^{-1}$)  & (mag)        & (rad)   & (mag)       & (rad)    & (mag)       & (rad) \\
\hline
$f_0$                 &2.176788       &0.5485(4)     &0.867(1) &0.4193(4)    &0.826(1)  &0.2603(4)    &0.681(1)\\
$2f_0$                &4.353576       &0.2531(4)     &3.930(2) &0.1993(3)    &3.936(2)  &0.1259(4)    &3.911(3)\\
...\\
$f_\mathrm{m}$        &0.014654       &0.0150(4)     &5.34(3)  &0.0132(4)    &5.38(3)   &0.0094(4)    &5.10(4)\\
$f_0+f_\mathrm{m}$    &2.191442       &0.0241(4)     &3.45(2)  &0.0168(4)    &3.42(2)   &0.0117(4)    &3.45(3)\\    
$2f_0+f_\mathrm{m}$   &4.368230       &0.0162(4)     &1.42(3)  &0.0142(4)    &1.36(3)   &0.0110(4)    &1.40(3)\\
...\\
$f_0-f_\mathrm{m}$    &2.162134       &0.0691(4)     &4.896(6) &0.0546(3)    &4.926(7)  &0.0347(4)    &5.02(1)\\
$2f_0-f_\mathrm{m}$   &4.338922       &0.0393(4)     &1.46(1)  &0.0313(4)    &1.47(1)   &0.0198(4)    &1.45(2)\\
...\\
$2f_0+2f_\mathrm{m}$  &4.382884       &0.0010(4)     &0.4(4)   &0.0017(4)    &1.4(2)    &0.0024(4)    &1.6(2)\\
$6f_0+2f_\mathrm{m}$  &13.09004       &0.0013(4)     &0.7(3)   &0.0016(3)    &5.8(2)    &0.0008(4)    &6.2(4)\\
...\\
$f_0-2f_\mathrm{m}$   &2.147480       &0.0028(4)     &4.8(1)   &0.0017(4)    &4.1(2)    &0.0015(4)    &4.4(3)\\
$2f_0-2f_\mathrm{m}$  &4.324267       &0.0058(4)     &0.13(7)  &0.0039(4)    &0.10(9)   &0.0025(4)    &6.2(1)\\
...\\
$6f_0-3f_\mathrm{m}$  &13.01676       &0.0009(4)     &5.9(4)   &0.0010(3)    &5.9(3)    &0.0006(4)    &6.1(6)\\
\\
$f_1'=f_2'-f_0$       &2.888023       &0.0033(4)     &3.9(1)   &0.0024(4)    &3.9(1)    &0.0016(4)    &4.2(2)\\
$f_2'$                &5.064811       &0.0016(4)     &0.7(3)   &0.0024(4)    &1.4(2)    &0.0023(4)    &1.4(2)\\
$f_3'=f_2'+f_0$       &7.241599       &0.0028(4)     &4.7(2)   &0.0020(4)    &4.8(2)    &0.0016(4)    &4.5(2)\\
$f_4'=f_2'+2f_0$      &9.418387       &0.0017(4)     &1.6(2)   &0.0014(4)    &1.6(2)    &0.0015(4)    &1.5(2)\\
\hline
\end{tabular}
\end{table*}

The times of the pulsation and Blazhko maxima, that is, the zero pulsation and modulation phases, are determined from the light-curve solutions. The epoch of the pulsation corresponds to the time of the maximum of the mean pulsation light curve, which is constructed as a synthetic light curve using the Fourier components of the pulsation frequency and its harmonics of the full light-curve solution. The epoch of the modulation is defined as the time when the amplitude of the fundamental frequency is the highest. This epoch does not necessarily coincide with the time of the highest amplitude phase of the modulation. The elements are given in the following equations:

\begin{equation}
 T_\mathrm{max}^\mathrm{XY\,And} = \mathrm{HJD}\,2\,454\,381.5513 + E^\mathrm{XY\,And}_\mathrm{puls} \cdot 0.3987247\,\mathrm{d}
\end{equation}

\begin{equation}
\label{eq:blephem_xy}
 T_\mathrm{Bl\,max}^\mathrm{XY\,And} = \mathrm{HJD}\,2\,454\,379.9 + E^\mathrm{XY\,And}_\mathrm{mod} \cdot 41.37\,\mathrm{d}
\end{equation}

\begin{equation}
 T_\mathrm{max}^\mathrm{UZ\,Vir} = \mathrm{HJD}\,2\,454\,470.2294 + E^\mathrm{UZ\,Vir}_\mathrm{puls} \cdot 0.4593925\,\mathrm{d}
\end{equation}

\begin{equation}
\label{eq:blephem_uz}
 T_\mathrm{Bl\,max}^\mathrm{UZ\,Vir} = \mathrm{HJD}\,2\,454\,472.9 + E^\mathrm{UZ\,Vir}_\mathrm{mod} \cdot 68.24\,\mathrm{d}
\end{equation}

\section{Comparison and details of the modulation of XY~And and UZ~Vir}

\begin{figure}
\centering
\includegraphics[width=8.7cm]{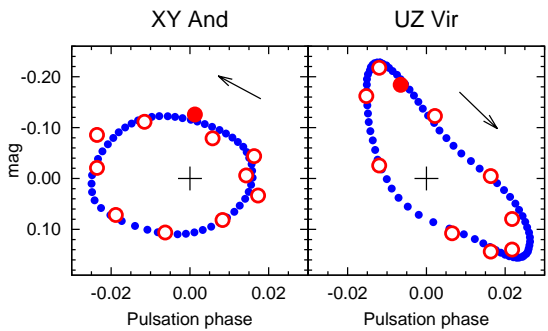}
\caption{Brightness and phase variations of the pulsation-light maxima of XY~And and UZ~Vir in $V$ band. Large (red) circles mark the observations in 11 and 10 Blazhko phase bins. Filled circles represent the bins nearest to zero Blazhko phase according to the elements given in Eqs.~\ref{eq:blephem_xy} and \ref{eq:blephem_uz}. Small (blue) dots are generated from synthetic light-curve solutions using the frequencies given in Tables~\ref{tbl:freqxya} and \ref{tbl:frequzv}. Relative phases and magnitudes of the maxima  with respect to the maximum of the mean pulsation light curve (marked by plus signs) are plotted. Arrows mark the directions of going around the loops; this is anti-clockwise for XY And and clockwise for UZ Vir.
\label{fig:tojas}}
\end{figure}

\begin{figure*}
\centering
\includegraphics[width=8.5cm]{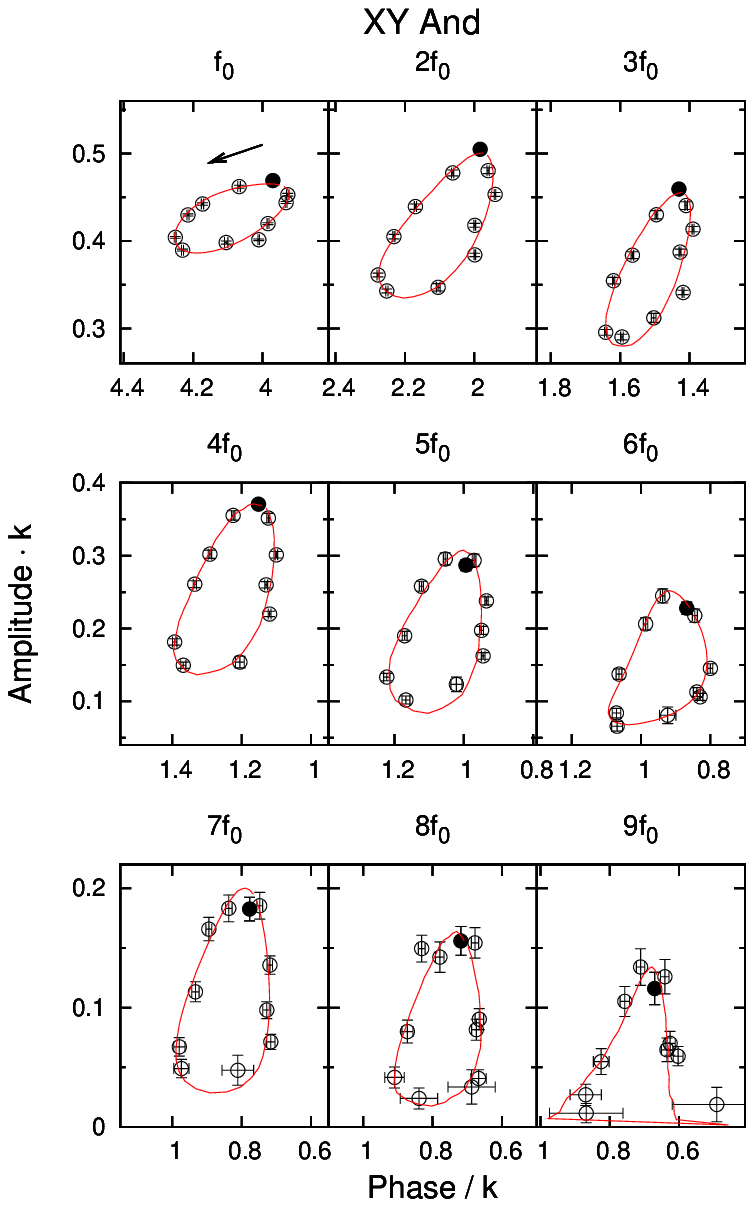}
\includegraphics[width=8.5cm]{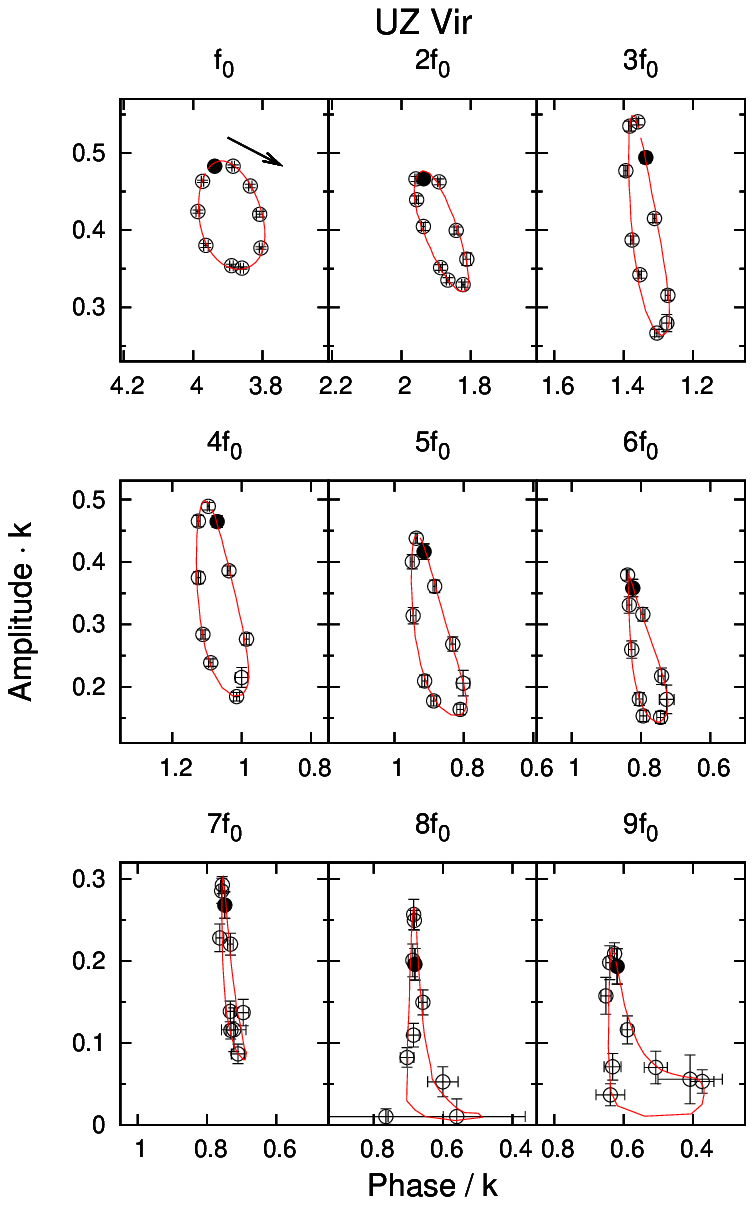}
\caption{The amplitude vs phase variations (with uncertainties) of the first nine harmonic orders of the pulsation light curves of XY~And and UZ~Vir during their modulation cycles. Data of the bins nearest to 0.0 Blazhko phase are marked with full dots. The solid lines show the same loops according to synthetic data. The amplitudes and phases are scaled by `$k$', the harmonic order, for the better visibility. Note that the phase range is the same in each panel: 0.6 rad. The horizontal axes are reversed in order to be in accordance with the variation of the phase of the maxima shown in Fig.~\ref{fig:tojas}. The directions of going around the curves are marked with arrows in the top-left panels. This direction remains the same for all the harmonic orders in both stars.
\label{fig:egg}}
\end{figure*}

\begin{figure*}
\centering
\includegraphics[width=7.8cm]{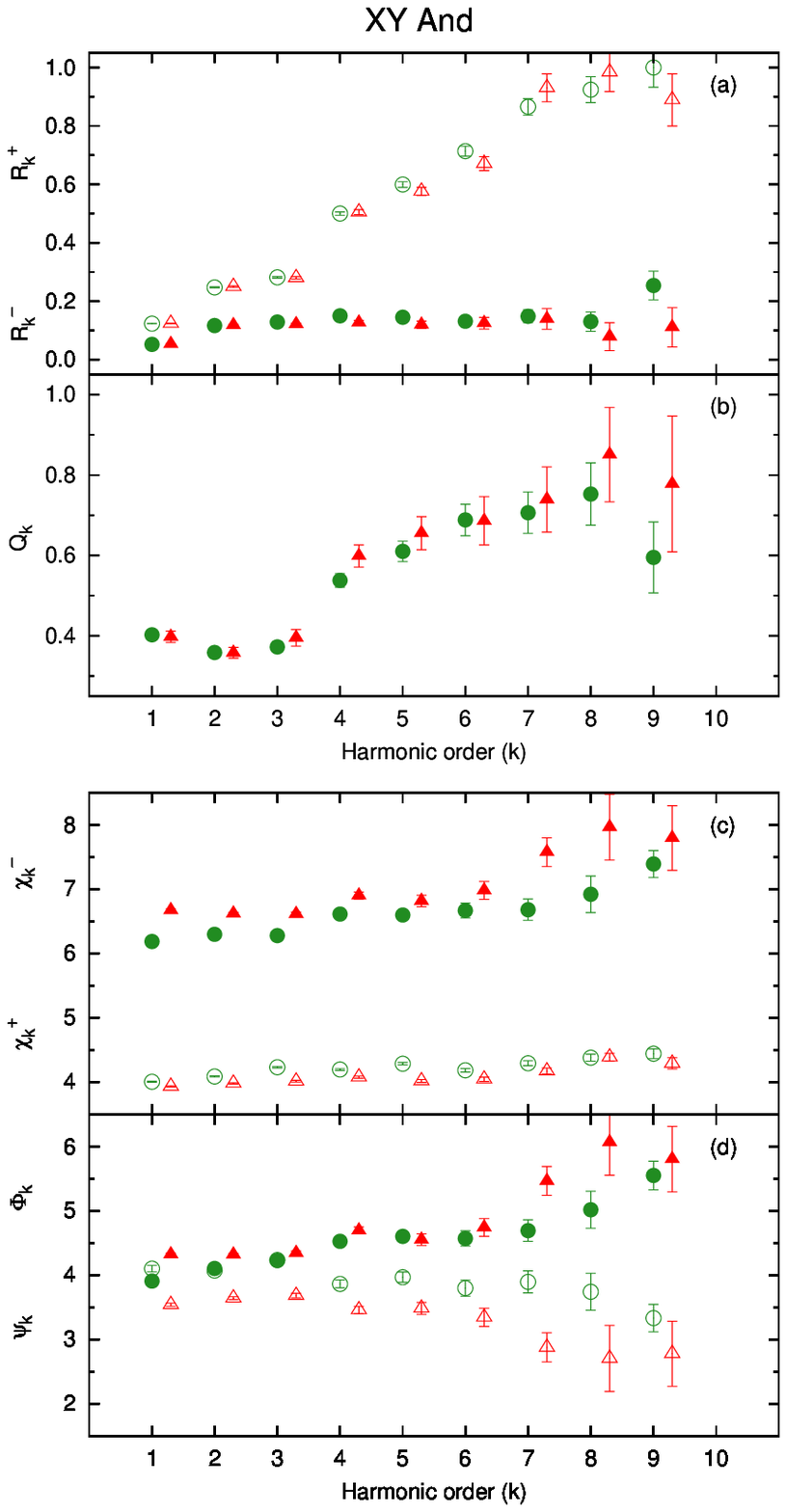}
\includegraphics[width=7.8cm]{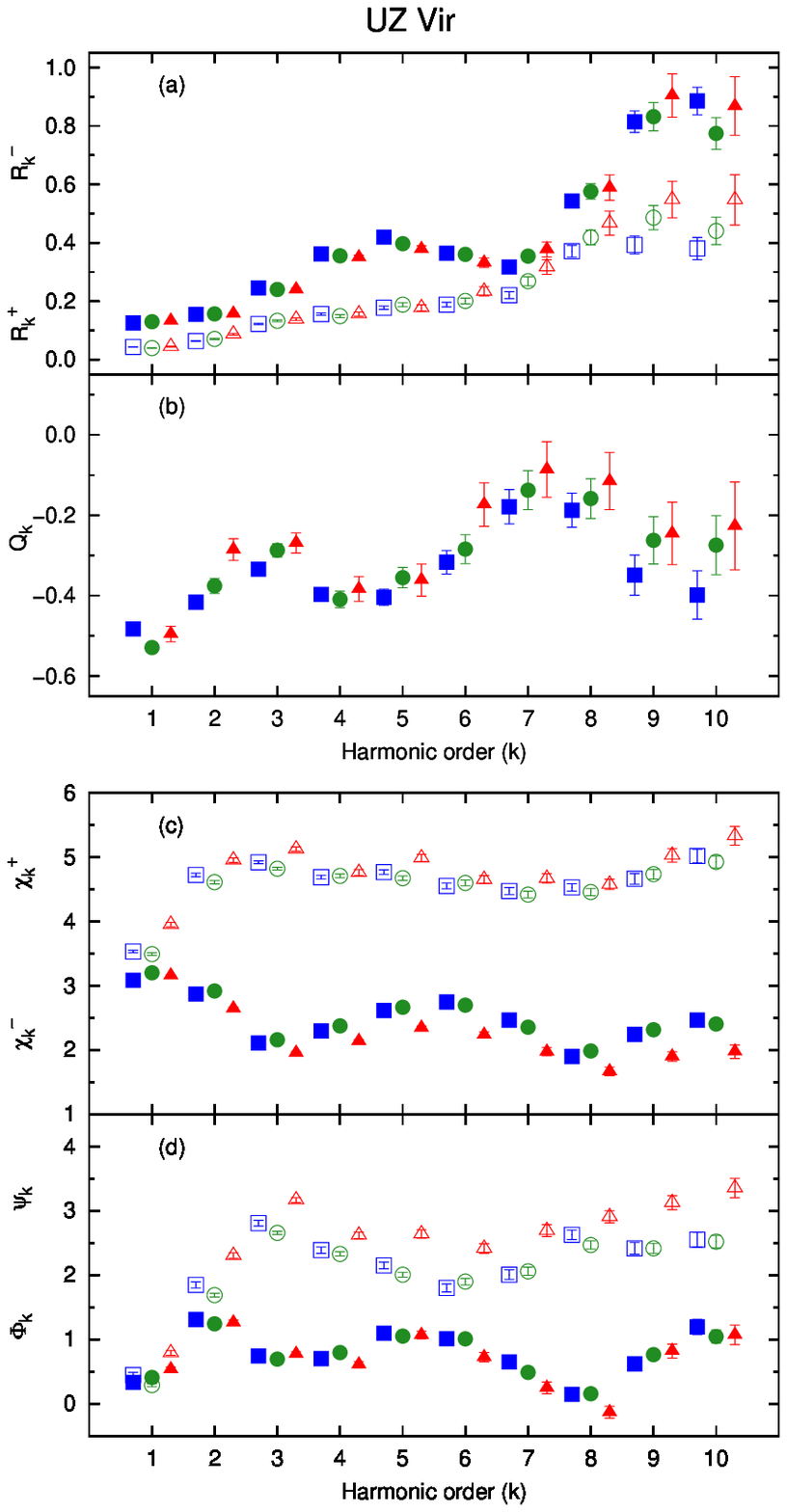}
\caption{Amplitude and phase relations of the triplet components of XY~And and UZ~Vir in the different harmonic orders. Squares (blue), circles (green) and triangles (red) denote the results in the $B$, $V$ and $I_{\mathrm{C}}$ bands, respectively. Results for the different photometric bands of the same harmonic order are shifted horizontally for a better visibility. From top to bottom: (a) $R^+_k=A_{kf_0 + f_\mathrm{m}}/A_{kf_0}$ (empty) and $R^-_k=A_{kf_0 - f_\mathrm{m}}/A_{kf_0}$ (filled) amplitude ratios of the triplet components relative to the harmonic components of the pulsation; (b) $Q_k=(A_{kf_0+f_\mathrm{m}}-A_{kf_0-f_\mathrm{m}})/(A_{kf_0+f_\mathrm{m}}+A_{kf_0-f_\mathrm{m}})$, the asymmetry parameter; (c) $\chi^+_k=\varphi_{kf_0 + f_\mathrm{m}}-\varphi_{kf_0}-\varphi_{f_\mathrm{m}}$ (empty) and $\chi^-_k=\varphi_{kf_0 - f_\mathrm{m}}-\varphi_{kf_0}+\varphi_{f_\mathrm{m}}$ (filled), the phase differences between the modulation components and the harmonics of the pulsation; (d) $\Phi_k=\varphi_{kf_0 + f_\mathrm{m}}+\varphi_{kf_0-f_\mathrm{m}}-2\varphi_{kf_0}$ (filled) and $\Psi_k=\varphi_{kf_0 + f_\mathrm{m}}-\varphi_{kf_0-f_\mathrm{m}}-2\varphi_{f_\mathrm{m}}$ (empty) phase relations of the pulsation and the modulation components.}
\label{fig:a-p}
\end{figure*}

As Fig.~\ref{fig:lc} shows, both the phase and the amplitude modulations of XY~And are pronounced, while the fix point on the rising branch of the light curve of UZ~Vir indicates that amplitude modulation is dominant in this star. The phase of the maximum brightness of both stars varies, however, within about the same, $\pm 0.02$ pulsation-phase range in both stars, as can be seen in Fig.~\ref{fig:tojas}. This contradictory result warns that the definition of the amplitude of the phase modulation is not at all straightforward.

The modulation of UZ~Vir is of a rare type, because the amplitude of the low-frequency modulation components of the triplets are much larger than the high-frequency ones. 
The amplitude difference between the two modulation components of a triplet is characterized by the asymmetry parameter: $Q = (A_+-A_-)/(A_++A_-)$, introduced by \cite{macho}.
Both the MACHO \citep{macho} and the Konkoly Blazhko Survey I and II \citep{aspc}  results show that the majority of the Blazhko stars have a positive asymmetry parameter.
The asymmetry parameter of UZ~Vir, calculated for the triplet around $f_0$ in $V$ band, has an extremely large negative value: $Q=-0.53\pm0.01$. Among the 731 Blazhko RRab stars in the MACHO LMC sample, only about 4 per cent have larger negative $Q$ values \citep[see fig. 10 of][]{macho} than this value.

As \cite{coast} pointed out, a connection exists between the sign of the $Q$ parameter and the traveling direction of the light maximum along the loop in the maximum phase\,--\,brightness plane. The negative $Q$ parameter of UZ~Vir reflects the property that its light maximum goes along the loop in clockwise direction (see Fig.~\ref{fig:tojas}), while the maximum of most of the Blazhko stars either goes in the opposite, counter-clockwise direction (as XY~And does), or the loop is degenerate.

Examining Fig.~\ref{fig:lc}, another significant difference between the modulation of the two Blazhko stars is discernible. The pulsation light curve of XY~And is much smoother in each Blazhko phase than that of UZ~Vir, especially around light minimum. The bumps and humps, characterizing the light variation of UZ~Vir, are completely missing from the light curve of XY~And. This difference is also reflected in the light-curve solutions; although the noise characteristics of the residual spectra of the two objects are similar, significantly more frequency components are needed (84) to describe the more complex brightness variation of UZ~Vir than for XY~And (57). The harmonic components of the pulsation are detected up to $25f_0$ in UZ~Vir, while only up to $13f_0$ in XY~And, which indicates that the mean pulsation light curve of XY~And is more simple than that of UZ~Vir, as well.

\subsection{Amplitude and phase relations of the pulsation light curve}

Fig.~\ref{fig:egg} shows the amplitude vs phase variations of the first nine harmonic orders of the pulsation light curves of XY~And and UZ~Vir in 11 and 10 phase bins of their modulation cycles, respectively. The solid lines show the same loops derived from synthetic light-curve solutions. The directions of going around the curves are marked with arrows in the top-left panels. This direction remains the same for all the harmonic orders in both stars.

The phase\,--\,amplitude loops are dissimilar in the different harmonic orders for both stars and none of the loops matches exactly the observed maximum phase\,--\,maximum brightness loops shown in Fig.~\ref{fig:tojas}. Furthermore, the maximum amplitude of the loops occurs in different Blazhko phases, as can be seen in Fig.~\ref{fig:egg}, where the
phases and amplitudes of the light curve at zero Blazhko phase are denoted by solid symbols. Some of the highest order loops show an interesting feature;  they seem to `open' at the lowest amplitudes in the 9th, and the 8\,--\,9th orders for XY~And and UZ~Vir, respectively. In accordance with its more complex light variation, the phase\,--\,amplitude loops of UZ~Vir show a large variety, from a circular shape (1st order) to a quasi-degenerate one (7th order), while, with the exception of the 9th order, all the phase\,--\,amplitude loops of XY~And are `egg'-shaped or elliptical.

\subsection{Amplitude and phase relations of the triplet components}

The relative amplitudes of the pulsation and modulation components, 
\begin{equation}
R^+_k=A_{kf_0 + f_\mathrm{m}}/A_{kf_0},
\end{equation}
\begin{equation}
R^-_k=A_{kf_0 - f_\mathrm{m}}/A_{kf_0}, 
\end{equation}
and the asymmetry parameter 
\begin{equation}
Q_k=(A_{kf_0+f_\mathrm{m}}-A_{kf_0-f_\mathrm{m}})/(A_{kf_0+f_\mathrm{m}}+A_{kf_0-f_\mathrm{m}})
\end{equation}
are derived for each harmonic order ($k$).

To follow the phase variations, the epoch-independent phase differences between the pulsation and the modulation components, 
\begin{equation}
\chi^+_k=\varphi^{}_{kf_0 + f_\mathrm{m}}-\varphi^{}_{kf_0}-\varphi_{f_\mathrm{m}}
\end{equation}
\begin{equation}
\chi^-_k=\varphi^{}_{kf_0 - f_\mathrm{m}}-\varphi^{}_{kf_0}+\varphi_{f_\mathrm{m}}
\end{equation}
and phase relations of the two modulation components, 
\begin{equation}
\Phi_k=\varphi^{}_{kf_0 + f_\mathrm{m}}+\varphi^{}_{kf_0-f_\mathrm{m}}-2\varphi^{}_{kf_0} 
\end{equation}
\begin{equation}
\Psi_k=\varphi^{}_{kf_0 + f_\mathrm{m}}-\varphi^{}_{kf_0-f_\mathrm{m}}-2\varphi^{}_{f_\mathrm{m}}
\end{equation}
are calculated according to the Fourier solutions of the light curves given in Tables~\ref{tbl:freqxya} and \ref{tbl:frequzv}.

Fig.~\ref{fig:a-p} shows the $R^+_k, R^-_k, Q^{}_k, \chi^+_k, \chi^-_k, \Phi^{}_k$, and  $\Psi^{}_k$ amplitude and phase relations of the triplet components in the different harmonic orders for the observed bands of the two studied stars. 

\subsubsection{XY~And}

Both the amplitude and phase variations show smooth, gradual, almost monotonic changes. The first panel (a) illustrates that while the low-frequency triplet components hardly changes as compared to the pulsation components, the amplitudes of the high-frequency components become as high as the amplitudes of the pulsation frequencies from the 7th harmonic order. As a result, the asymmetry parameter, $Q^{}_k$, is increasing towards higher harmonic orders (panel b).

The bottom set of panels displays the phase-difference variations of the triplet components. Each of the four phase relations shows quite a regular behaviour. It is interesting to note that while only 0.5 rad gradual increase of the phase difference between the higher amplitude triplets and the pulsation components ($\chi^+_k$) is observed, the phase difference of the lower amplitude triplet component ($\chi^-_k$) is less stable, it changes within about 1.5 rad. The $\Phi^{}_k$ and $\Psi^{}_k$ phase relations also show monotonic changes with increasing gradients towards higher harmonic orders.

The most interesting feature of the phase relations is that each of the four shows a definite colour dependence, which is the most pronounced for $\chi^-_k$ and $\Psi^{}_k$. These phase relations in the $V$ and $I_\mathrm{C}$ bands differ systematically in each harmonic order by 0.3\,--\,1.0 rad. The $Q^{}_k$ asymmetry parameter may also be somewhat colour dependent; the asymmetry of the triplets seems to be slightly larger in the $I_C$ band than in the $V$ band. This tendency, however, is not really significant.

\subsubsection{UZ~Vir}

The amplitude and phase relations of the triplets of UZ~Vir (right-hand panels of Fig.~\ref{fig:a-p}) show a more complex behaviour than for XY~And. Most probably, this is the consequence of the appearance and disappearance of the bump/hump features on the light curve during the modulation cycle.

The amplitude variation of the lower amplitude triplet component, $R^+_k$, is smoother than that of the higher amplitude triplet component, $R^-_k$, which shows local maxima at the 5th and the 9\,--\,10th orders. The variation of the asymmetry parameter is also complex, it has local minima (taking into account the negative values of $Q^{}_k$, these correspond to large asymmetry) in the first, the 4\,--\,5th and 9\,--\,10th harmonic orders. Towards higher orders, the asymmetry of the triplets ($|Q^{}_k|$) shows a decreasing tendency, while it increases for XY~And.

The phase behaviour of the triplet components of UZ~Vir is not smooth either. The total ranges of the variations of $\chi^+_k$, $\chi^-_k$ and $\Phi^{}_k$ are 1.5\,--\,2 rad, while the variation of $\Psi^{}_k$ extends to 2.5 rad. The $\chi^-_k$ phase difference follows similar changes to the variations of the asymmetry parameter with local minima in the 3rd and 8th orders. With the exception of the first order of $\Phi^{}_k$, the variations of $\Psi^{}_k$ and $\Phi^{}_k$ reflect the changes of $\chi^+_k$ and $\chi^-_k$, respectively.

Colour dependence of the phase relations of UZ~Vir is also detected. The $\chi^+_k$ and $\Psi^{}_k$ values are systematically larger by 0.1\,--\,0.5 rad in the $I_C$ band than in the $B$ and $V$ bands, while the colour dependence seen in the  $\chi^-_k$ data is the opposite.  With the exception of the first order, the asymmetry parameter of UZ~Vir is also colour dependent. Contrary to the results for XY~And, the triplets are less asymmetric towards longer wavelengths here.

\section{Physical parameters and their variation during the modulation}

The variations of the pulsation-averaged atmospheric parameters of XY~And and UZ~Vir during the Blazhko cycle are derived using the inverse photometric Baade--Wesselink method \citep[IPM;][]{ip}. These parameters are the effective temperature ($T_\mathrm{eff}$), luminosity ($L$), radius ($R$) and effective surface gravity ($\log g$). Utilizing multicolour light curves and  synthetic colours from static atmosphere models \citep{cast}, the method finds the best fitting model parameters through the variation of template $T_{\mathrm {eff}}$ and $V_{\mathrm {rad}}$ curves. In order to test the stability and reliability of the method, different $V_{\mathrm {rad}}$ template curves and different weightings are applied. Therefore, similarly to the previous analyses, the IPM code is run with four different internal settings \citep[for details see][Table~1]{ip} to estimate the inherent, method-specific uncertainty of the results.

\subsection{Constant parameters}
\label{sect:stpar}

\begin{table*}
\centering
\caption{Mean physical parameters and their uncertainties derived from the mean pulsation light curves of XY~And and UZ~Vir with the inverse photometric method (IPM) using static atmosphere models with different [Fe/H] metallicities. The luminosities, temperatures and masses are constrained to match the evolutionary possible values for horizontal-branch stars. The mean physical parameters are also required to fit the pulsation period via the pulsation equation. $(M_V)$ denotes the intensity-averaged absolute $V$ magnitude.
\label{tbl:ipm_totalmean}
}
\begin{tabular}{lccccccccc}
\hline
Object & [Fe/H] & $\mathfrak{M}$       & $d$         & $(M_V)$         & $R$           & $L$          & $T_\mathrm{eff}$ \\ 
       & (dex)  &($\mathfrak{M}_\odot$)& (pc)        & (mag)           & ($R_\odot$)   & ($L_\odot$)  &        (K)       \\
\hline
XY And & $-0.68$& $0.58\pm0.01$        & $3700\pm200$& $0.76\pm 0.05$  & $4.15\pm 0.20$& $40.5\pm 2.5$ & $7100\pm 100$     \\[1mm]
UZ Vir & $-0.65$& $0.61\pm0.01$        & $2850\pm150$& $0.77\pm 0.05$  & $4.54\pm 0.20$& $41.0\pm 2.5$ & $6800\pm 100$     \\
UZ Vir & $-0.90$& $0.59\pm0.01$        & $2900\pm150$& $0.71\pm 0.05$  & $4.55\pm 0.20$& $43.5\pm 2.5$ & $6900\pm 100$     \\
UZ Vir & $-1.26$& $0.57\pm0.01$        & $2950\pm150$& $0.65\pm 0.05$  & $4.56\pm 0.20$& $46.0\pm 2.5$ & $7000\pm 100$     \\
\hline
\end{tabular}
\end{table*}

To apply the IPM, the Blazhko-phase-independent parameters have to be known first. These are the metallicity ([Fe/H]), mass ($\mathfrak{M}$), distance ($d$), and the mean dereddened colours of the variables averaged over both the pulsation and modulation periods.

The metallicity of XY~And was measured by spectroscopic \citep{layden} and photometric \citep{jk96} means. Both methods yield the same, $-0.68$\,dex result \citep[][table~5]{kbs1}. Therefore, this  value, as a reliable estimate for the metallicity of XY~And, is accepted and applied throughout the following analysis. No spectroscopic measurement of the metallicity of UZ~Vir has been published. It was shown by \cite{kbs1} that a well defined mean pulsation light curve of a Blazhko star yields a reliable metallicity value according to the [Fe/H]$(P,\Phi_{31})$ formula \citep{jk96} even for strongly modulated RR~Lyrae stars. This method gives [Fe/H]\,$=-0.90$\,dex for UZ~Vir. However, the example of RZ~Lyr warns that, in certain cases, the photometric metallicity may differ by as large as $\sim0.4$\,dex from the spectroscopic value \citep{rzl}. Therefore, the effect of the choice of the input metallicity on the IPM results is checked by calculating the mean physical parameters of UZ~Vir using two other metallicity values, as well. As only evolutionary possible values of the mean physical parameters are accepted, the other metallicity values ($-0.65$ and $-1.26$) were selected from the [Fe/H] values of the horizontal branch evolutionary models of \cite{dorman92}.

Our differential photometric observations do not allow the determination of the mean (reddened) colours of these two variables, since no standard multicolour magnitude measurement of the comparison stars or of any other object in the field of view is available. \cite{schmidt} give the intensity-averaged values of $(V)-(R) = 0.22\pm0.02$ and $0.24\pm0.03$\,mag for XY~And and UZ~Vir, respectively. However, these colour indices should be treated with caution, since they are calculated only from 10\,--\,20 data points per band, observed in different Blazhko phases. Therefore, the uncertainties of these colours are estimated to be a few hundredth magnitude higher than the values given by \cite{schmidt}. According to the extinction maps of \cite{schlegel}, the interstellar reddening [$E(B-V)$] towards the directions of XY~And and UZ~Vir are $0.050$ and $0.026$~mag. Since these values also have uncertainties on the order of 0.01~mag, and because these are only upper limits for the actual reddenings, the possible colour ranges, consequently the possible overall temperature ranges, are too wide to impose any meaningful constraint on the IPM fitting process.

Therefore, the temperature, luminosity and mass of both stars have been constrained using the pulsation equation and horizontal-branch (HB) evolutionary-model results. According to the pulsation equation of \citet[][eq. 1a]{marconi} and the HB evolutionary models of \citet[][figs.~7\,--\,10]{dorman92}, the pulsation periods of the studied objects correspond to the following physical parameter ranges: 7000\,--\,7200\,K and 6700\,--\,7100\,K temperature, 38\,--\,43\,$L_\odot$ and 38.5\,--\,48.5\,$L_\odot$ luminosity for XY~And and UZ~Vir, respectively.

Applying these constraints, the IPM is run for the mean pulsation light curves of both objects to determine consistent sets of the mean and constant parameters. The results from these runs are listed in Table~\ref{tbl:ipm_totalmean}. To check the effect of the uncertainty in the metallicity of UZ~Vir, the mean physical parameters are calculated for three possible metallicity values. The distances are estimated accepting $\langle V \rangle = 13.73$ and 13.05 mag \citep{schmidt} and assuming $A(V) = 3.14 E(B-V) = 0.157$ and 0.082 interstellar absorption for XY~And and UZ~Vir, respectively. The uncertainties given in Table~\ref{tbl:ipm_totalmean} are estimated by taking into account the inherent error of the method, the uncertainties of the input parameters and the possible ranges of the constrained parameters according to the HB evolutionary models of \cite{dorman92}.  The data given in Table~\ref{tbl:ipm_totalmean} prove that the three possible metallicity values of UZ Vir results in similar uncertainties of the output parameters as the other factors together. The only exception is the radius, which is practically the same for all three input metallicities.

\label{sect:varphyspar}
\begin{figure*}
\centering
\hskip7mm
\includegraphics[width=8.3cm]{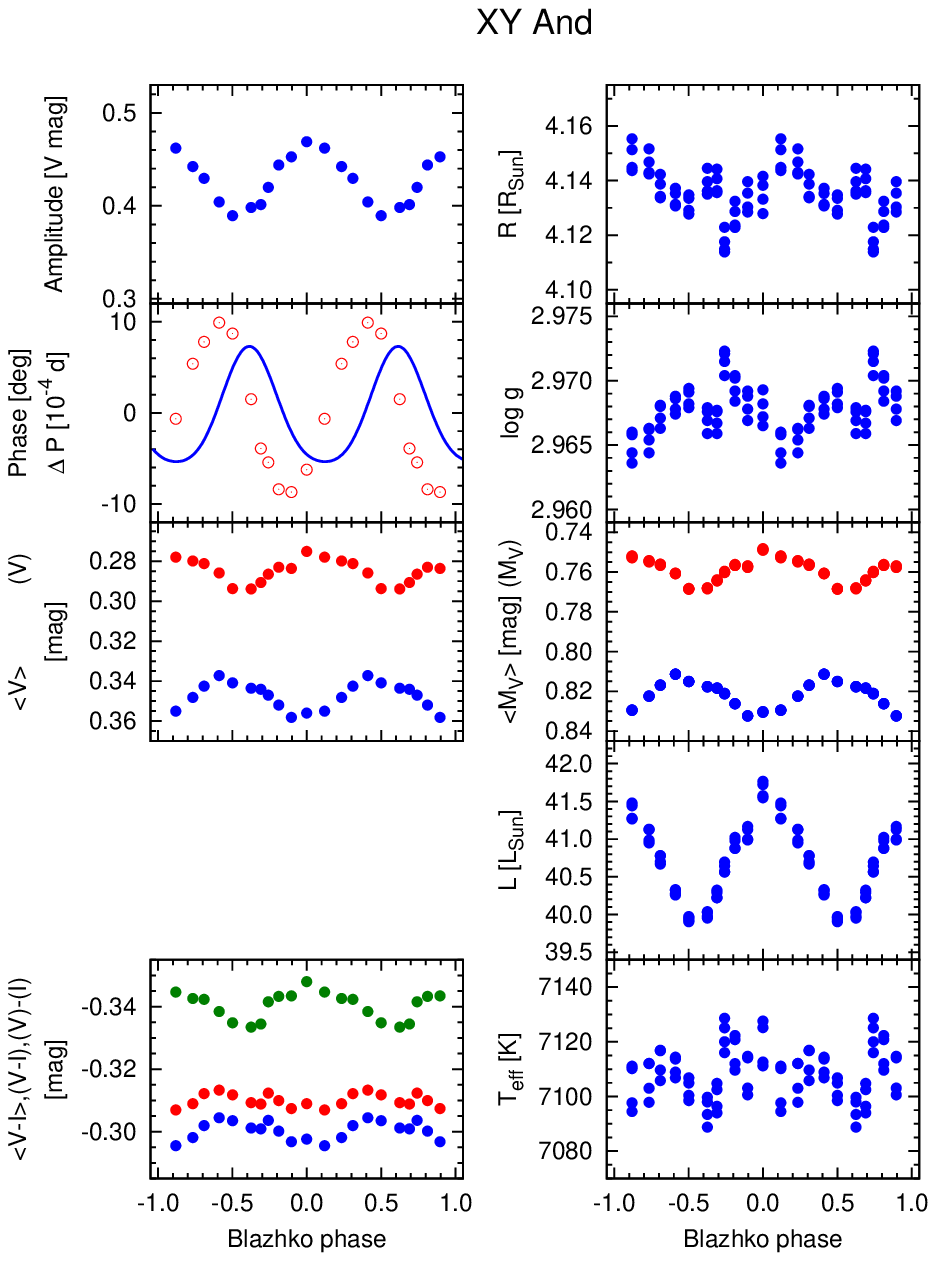}
\hskip2mm
\includegraphics[width=8.3cm]{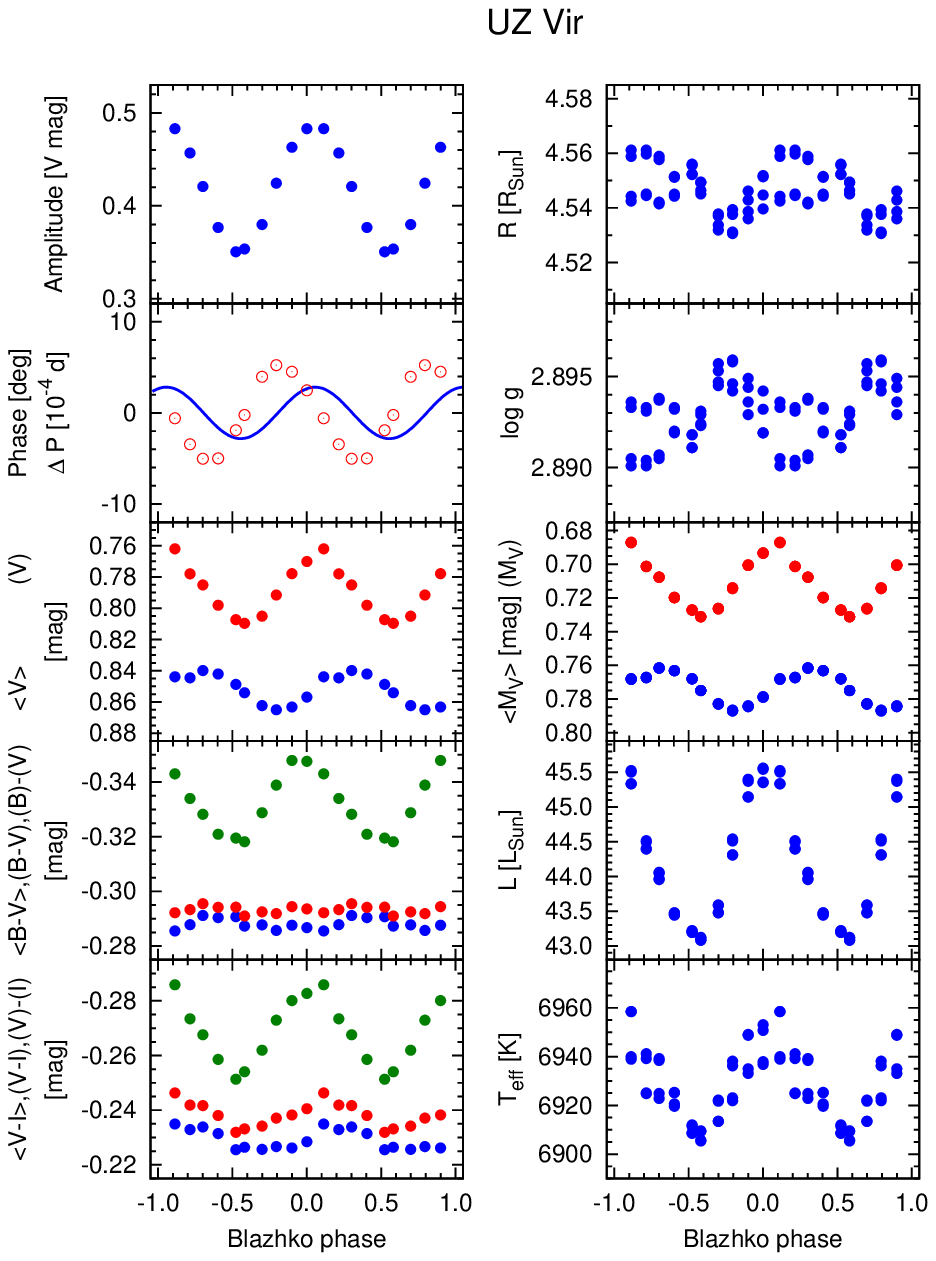}
\caption{The mean observed (left-hand panels) and the IPM-derived physical parameters (right-hand panels) of XY~And and UZ~Vir in different phases of the modulation. From top to bottom, the left-hand panels show: amplitude modulation -- amplitude of the $f_0$ frequency component; pulsation-phase or period modulation -- variation of the phase of the $f_0$ pulsation component (empty circles) and deviation of the instantaneous pulsation period from the mean pulsation period (continuous line); pulsation-averaged $V$ magnitude; pulsation-averaged $B-V$ colour; pulsation-averaged $V-I_\mathrm{C}$ colour. Magnitude- and intensity-averaged brightnesses and colours are denoted by angle and round brackets, respectively. From top to bottom, the right-hand panels show the pulsation averages of the following physical parameters: radius, surface gravity, absolute visual magnitude, luminosity, and effective temperature. Results of four different settings of the IPM are shown in order to indicate the inherent uncertainity of the method.
\label{fig:ipres}
}
\end{figure*}

\subsection{Blazhko-phase-dependent parameters}

Fixing the Blazhko-phase-independent physical parameters to their values determined in Sect.~\ref{sect:stpar} and listed in Table~\ref{tbl:ipm_totalmean}, the IPM is run using the light curves of XY~And and UZ~Vir in 11 and 10 different, disjunct phase bins of the modulation, respectively. In order to obtain the highest possible number of Blazhko phase bins with full pulsation-phase coverage, the bins are not exactly equidistant, but spread over the Blazhko cycle almost evenly. Each bin is narrow enough for the pulsation light curves not to be affected by Blazhko-phase smearing.

The IPM is run for UZ Vir with the three possible metallicities listed in Table~\ref{tbl:ipm_totalmean} to check its effect on the results. It is found that the different metallicity values affect only the means of the physical parameters, but the amplitudes and the phase relations of their variations with Blazhko phase remains practically the same. Therefore, only the results obtained with the most likely value of [Fe/H]\,$=-0.90$\,dex are shown.

The IPM gives the radius, luminosity, temperature and surface-gravity changes during the pulsation in each phase bin of the modulation. The pulsation-averaged mean values of these quantities characterize the global mean changes in the stellar parameters during the modulation cycle. Similarly to the IPM results of the previous investigations of Blazhko stars, the actual values of the fixed or constrained physical parameters influence the averages of the other physical parameters, nevertheless, they affect the variations during the Blazhko cycle only slightly. The amplitudes of the solutions differ somewhat if the input parameters are varied within their possible ranges (Table~\ref{tbl:ipm_totalmean}), but the relative sign of their variations remains unchanged.

The appearance of $f_m$ in the frequency solution of the light curves reflects the variation of magnitude averaged mean brightness in the different bands during the modulation. The IP method, utilizing  colour information as well, discloses what luminosity, radius and temperature changes are required to describe the observed changes in the mean magnitudes. The IPM results on XY~And and UZ~Vir are plotted against the modulation phase in Fig.~\ref{fig:ipres}. Here, the observed (left-hand panels) and IPM-derived (right-hand panels) variations of the pulsation-averaged parameters during the Blazhko modulation are plotted. The observed parameters (from top to bottom) are the amplitude of the main pulsation frequency ($f_0$); the $\Phi_{1}$ phase variation of this frequency component and the pulsation-period variations deduced from the observed phase variations of the $f_0$ component; and the different averages of the light and colour curves according to magnitude- and intensity-scale representations.

\section{Discussion and Summary}

\subsection{Light-curve modulation}

The Fourier spectra of the light curves of XY~And and UZ~Vir are described by multiplet frequencies ($kf_0\pm nf_\mathrm{m}$) and the modulation frequency ($f_\mathrm{m}$), typical of Blazhko stars. The multiplet structures are triplets and quintuplets, with a single septuplet component in the spectrum of UZ~Vir ($6f_0-3f_\mathrm{m}$). The phase of the rising branch of XY~And is strongly modulated, while a fix point on the rising branch of UZ~Vir (see Fig.~\ref{fig:lc}) indicates that amplitude modulation dominates its light-cure variation. Looking at the phase variations of the maximum brightness of the two stars, similar-amplitude changes are observed as shown in Fig.~\ref{fig:tojas}. This controversial behaviour of the phase variations of the rising branches and the maxima warns that the strength of the phase modulation at different parts of the pulsation light curve can be rather different.

The pulsation light curve of XY~And is much smoother in each Blazhko phase than that of UZ~Vir. Consequently, the light-curve solution of XY~And involves less pulsation and modulation components than that of UZ~Vir.

The modulation of UZ~Vir is of a rare type, since the amplitude of the low-frequency triplet components are significantly larger than that of the high-frequency ones. Accordingly, the light maximum goes in clockwise direction around the loop on the pulsation phase\,--\,brightness plane (Fig.~\ref{fig:tojas}), while the maximum of most of the Blazhko stars goes around in the opposite direction, or the loop is degenerate.

For the first time, the amplitude and phase relations of the triplet components are also studied in detail. No common feature of the variations of the defined parameters of the two stars has been recognized. This result supports that Blazhko stars behave very individually. It is difficult, if at all possible, to find any overall, common property of the modulation. The most interesting finding of this investigation is the detection of the colour dependency of the asymmetry parameter and the phase relations. The colour dependence of these quantities is the opposite of each other in the two stars, which might be connected to their opposite-sign amplitude and phase variations during the Blazhko cycle. The colour dependence of the amplitude and phase relations of the triplets of other Blazhko stars has to be studied to decide whether such a connection really holds. As the detected colour dependence carries important physical information on the modulation, its further investigation and explanation should be an important step in the description and interpretation of the Blazhko effect.

\subsection{Additional frequencies}

The additional frequencies in RR~Lyrae stars have been discovered only recently, as very precise and extended photometry is needed to find these low-amplitude signals.

Both XY~And and UZ~Vir show variability besides the pulsation and modulation with an additional, independent frequency. Series of four peaks appear in the Fourier spectra of both objects with separation that equals to their pulsation frequency. One of these peaks is identified as a linearly independent frequency component, while the other three members of the series are interpreted as linear combinations of this and the pulsation frequency. The frequency ratios of the possible main components of the additional frequency series ($f_0/f$) are 0.325 for XY And and 0.754, 0.430 or 0.301 for UZ Vir. The 0.301, 0.325 and 0.430 frequency ratio values cover the possible regime of the  third--sixth radial overtones. Unfortunately, the actual frequencies of the higher order radial modes on a large grid of stellar parameters have never been published. Thus we do not know whether or not the observed frequency ratios fit any higher order radial mode, indeed. However, as RR Lyrae models indicate that these higher order modes are strongly damped, therefore we consider it unlikely that radial modes were detected in these stars.

In the case of XY~And, two possible explanations of the additional-frequency series have been raised: (a) one frequency of the series might be a non-radial mode, (b) all four frequencies correspond to a secondary modulation of an extremely short, $\sim5.01$\,d period. For explaining the series of additional frequencies in UZ~Vir, we also raised two possibilities: (a) the frequency at 2.89\,cd$^{-1}$ might be the first radial overtone, however, with an unusually large $f_0/f_1$ ratio of 0.754, (b) one of the two highest amplitude components of the additional frequency series corresponds to a non-radial mode. The recent detection of an additional frequency with 0.753-758 frequency ratio in the {\it Kepler} data of RR Lyr (Moln\'ar et al. submitted to ApJ), which has been identified with the first overtone mode, favours the first possibility in the case of UZ Vir, too.

It is of interest to examine the reported occurrences of other, additional frequencies among RR Lyrae variables. Both double mode RR Lyrae stars with extended satellite observations \citep[AQ Leo and CoRoT ID 0101368812][]{aql,chadid} show  additional frequencies beyond the fundamental and the first overtone modes. Many Blazhko variables have one or more extra, independent frequencies in their light variation in addition to the pulsation and modulation components. Such frequencies have been reported by \citet{v1127aql}, \citet{poretti} and \citet{gugg,gugg12}, see also Table~2. of \citet{benko}. The $\pm 12.5 f_\mathrm{m}$ peaks around some of the pulsation components in MW~Lyr \citep{mw1} might also be connected to an additional frequency, as suggested by \citet{poretti}. In contrast, only two of the 19 non-modulated RRab stars observed by $Kepler$ show frequencies besides the fundamental-mode pulsation \citep{nemec}, and no additional frequency has been detected in the only non-Blazhko $CoRoT$ RRab star analysed so far \citep{paparo}. In the residual spectra of the 14 non-modulated RRab stars observed in the Konkoly Blazhko Survey I \citep[see references in][]{kbs1}, no additional frequency has been found either. Although the majority of the $Kepler$ Blazhko targets still lack thorough analysis, it is clear that the occurrence rate of the appearance of additional frequencies is much higher among modulated RRab stars than in non-modulated ones. A possible explanation might be that the changing physical parameters during the Blazhko-cycle \citep{mw2} temporarily fulfill the conditions required for the excitation of the other observed (radial or non-radial) modes, whereas the chance to satisfy the excitation conditions  is much smaller for stable-light-curve RRab stars. A similar explanation has been suggested for the appearance of a peculiar bump on the descending branch of RZ Lyr in the low-amplitude phases of its modulation \citep{rzl}. However, the excited, additional mode of RZ Lyr is supposed to be in resonance with one of the harmonics of the main pulsation frequency, which explains why it does not appear in the Fourier spectrum as an additional frequency directly.

\subsection{Physical-parameter changes during the modulation}

The modulational changes in the atmospheric parameters of XY~And and UZ~Vir have been derived applying the IP method. Contrary to the many differences between the Blazhko-modulation properties of the two objects, their mean physical parameters are rather similar; XY~And is only $\sim$150\,K hotter, $\sim$4\,$L_\odot$ fainter, $\sim$0.4\,$R_\odot$ smaller, $\sim$0.01\,$\mathfrak{M}_\odot$ less massive and $\sim$0.22\,dex more metal rich, than UZ~Vir. The changes in the mean temperature, radius and luminosity during the Blazhko cycle of the two stars do not show any significant difference either.

Both objects show strong luminosity variation during the modulation: $\Delta L \approx2\,L_\odot$ and $\approx2.5\,L_\odot$ are derived for  XY~And and UZ~Vir, respectively. Similarly to all Blazhko RR Lyrae stars analysed with the IPM so far, XY~And and UZ~Vir are the brightest when the pulsation amplitude is the highest.

The radius variation is weak; the average radius is the highest around Blazhko maximum in both variables. The IPM results show only marginal temperature variation of XY~And during the Blazhko cycle, but the $T_\mathrm{eff}$ and $(V)-(I_\mathrm{C})$ curves together suggest that XY~And is about 10\,--\,20\,K warmer at the high-amplitude phase of the modulation than when the pulsation amplitude is the lowest. The IPM results on UZ~Vir indicate only a modest but definite modulational variation of about 30\,K in temperature. Both the radius and temperature are the highest at Blazhko maximum both in XY~And and in UZ~Vir.

The Blazhko model proposed by \cite{stothers} postulates a changing magnetic field in the envelope of the star, which influences the parameters of convection. The model makes predictions about the phase relation between the amplitude and period modulations and between the period modulation and radius variation \citep{stothers2011}, namely, these relations depend on the temperature of the star. For hot RRab stars, the \cite{stothers} model predicts that the pulsation period is decreasing at high-amplitude Blazhko phase (the period and amplitude variations are in anti-phase), while for stars below the crossover temperature of about $6400$\,K, the period is increasing at the same Blazhko phase (the changes of the two quantities are in phase). These relations are shown in the top two panels in the left columns of Fig.~\ref{fig:ipres} for the two studied objects (the pulsation-period changes are plotted with continuous line). The two variables have similar temperatures well above the predicted crossover value, and they also have similar pulsation periods and metallicities. Still, the phase relations between their pulsation amplitude and pulsation period variation are the opposite of each other. This behaviour is also demonstrated in Fig.~\ref{fig:tojas}, which shows that maxima of the two objects go along their modulational loops in the pulsation phase\,--\,brightness plane in opposite direction.

The Stothers model also expects a relation between the pulsation period and radius variation during the Blazhko cycle \citep{stothers2011}. The model predicts anti-correlated variation between these two quantities in hot RRab stars, which turns into correlation for cooler RRab variables. However, both objects are hot, short-period RRab stars, and both have the highest radius at Blazhko maximum, while their pulsation-period variations are the opposite of each other.

These findings are not in accord with the predictions of the Stothers model \citep{stothers,stothers2011}, which claims that two Blazhko stars with such similar properties are expected to display similar relations between their amplitude, pulsation period and radius variations. Our results show that even if the fundamental idea of the Stothers model is right, the relations between the studied properties of the Blazhko stars are more complicated than predicted by the model in its present form.

\section*{Acknowledgments}

The support of OTKA grants K-076816 and K-081373 is acknowledged.

\appendix

\section{Samples of the electronic tables}
\label{appendix:electronic}

The full tables, containing the observations of UZ~Vir and XY~And (Tables~\ref{tbl:dataxyav}\,--\,\ref{tbl:datauzvi}), are available as supporting information.

\begin{table}
\centering
\caption{CCD $\Delta V$ observations and derived colour indices of XY~And, relative to the comparison star
2MASS 01270016+3404307.
 \label{tbl:dataxyav}
}
  \begin{tabular}{ccc}
  \hline
HJD\,$-$\,2400000  &$\Delta V$(mag) &$\Delta V-I$(mag)\\
\hline
54390.46709  &0.466           &-0.211\\
54390.47001  &0.483           &-0.201\\
...          &...             &...\\
\hline
\end{tabular}
\end{table}

\begin{table}
\centering
\caption{CCD $\Delta I_\mathrm{C}$ time series of XY~And.
 \label{tbl:dataxyai}
}
  \begin{tabular}{cc}
  \hline
HJD\,$-$\,2400000  &$\Delta I_\mathrm{C}$(mag)\\
\hline
54390.46855  &0.677\\
54390.47148  &0.678\\
...          &...\\
\hline
\end{tabular}
\end{table}

\begin{table}
\centering
\caption{CCD $\Delta V$ observations and derived colour indices of UZ~Vir, relative to the comparison star
2MASS 13082756+1322403.
 \label{tbl:datauzvv}
}
  \begin{tabular}{cccc}
  \hline
HJD\,$-$\,2400000  &$\Delta V$(mag) &$\Delta B-V$(mag) &$\Delta V-I$(mag)\\
\hline
54504.57397  &1.183           &-0.210            &-0.109\\
54504.57976  &1.191           &-0.217            &-0.102\\
...          &...             &...               &...\\
\hline
\end{tabular}
\end{table}

\begin{table}
\centering
\caption{CCD $\Delta B$ time series of UZ~Vir.
 \label{tbl:datauzvb}
}
  \begin{tabular}{cc}
  \hline
HJD\,$-$\,2400000  &$\Delta B$(mag)\\
\hline
54504.57158  &0.975\\
54504.57737  &0.986\\
...          &...\\
\hline
\end{tabular}
\end{table}

\begin{table}
\centering
\caption{CCD $\Delta I_\mathrm{C}$ time series of UZ~Vir.
 \label{tbl:datauzvi}
}
  \begin{tabular}{cc}
  \hline
HJD\,$-$\,2400000  &$\Delta I_\mathrm{C}$(mag)\\
\hline
54504.57515  &1.309\\
54504.58108  &1.296\\
...          &...\\
\hline
\end{tabular}
\end{table}

\label{lastpage}

\begin{thebibliography}{}


\bibitem[\protect\citeauthoryear{Alcock et al.}{2003}]{macho}        Alcock C. et al., 2003, ApJ, 598, 597
\bibitem[\protect\citeauthoryear{Benk\H{o} et al.}{2010}]{benko}     Benk\H{o} J. M. et al., 2010, MNRAS, 409, 1585
\bibitem[\protect\citeauthoryear{Buchler \& Koll\'ath}{2001}]{bk01}    Buchler R., Koll\'ath Z. 2001, ApJ, 555, 961
\bibitem[\protect\citeauthoryear{Buchler \& Koll\'ath}{2012}]{bk12}    Buchler R., Koll\'ath Z. 2012, ApJ, 731, 24
\bibitem[\protect\citeauthoryear{Castelli \& Kurucz}{2003}]{cast}    Castelli F., Kurucz R. L., 2003, in Piskunov N., Weiss W. W., Gray D. F., eds, Proc. IAU Symp. 210, Modelling of Stellar Atmospheres. Astron. Soc. Pac., San Francisco, p. A20
\bibitem[\protect\citeauthoryear{Dziembowski \& Cassisi}{1999}]{dc} Dziembowski W., Cassisi S. 1999, AcA, 49, 371
\bibitem[\protect\citeauthoryear{Chadid et al.}{2010}]{v1127aql}     Chadid M. et al., 2010, A\&A, 510, 39
\bibitem[\protect\citeauthoryear{Chadid}{2012}]{chadid} Chadid, M. 2012, A\&A, 540, 68
\bibitem[\protect\citeauthoryear{Deeming}{1975}]{de75}               Deeming T. J., 1975, Ap\&SS, 36, 137
\bibitem[\protect\citeauthoryear{Dorman}{1992}]{dorman92}            Dorman B., 1992, ApJS, 81, 221
\bibitem[\protect\citeauthoryear{Gruberbauer et al.}{2007}]{aql} Gruberbauer M. et al., 2007, MNRAS, 379, 1498
\bibitem[\protect\citeauthoryear{Guggenberger et al.}{2011}]{gugg}   Guggenberger E., Kolenberg K., Chapellier E., Poretti E., Szab\'o R., Benk\H{o} J. M., Papar\'o M., 2011, MNRAS, 415, 1577
\bibitem[\protect\citeauthoryear{Guggenberger et al.}{2012}]{gugg12}   Guggenberger E., et al.,  2012, MNRAS, 424, 649
\bibitem[\protect\citeauthoryear{Jurcsik \& Kov\'acs}{1996}]{jk96}   Jurcsik J., Kov\'acs G., 1996, A\&A, 312, 111
\bibitem[\protect\citeauthoryear{Jurcsik et al.}{2008a}]{mw1}        Jurcsik J. et al., 2008a, MNRAS, 391, 164
\bibitem[\protect\citeauthoryear{Jurcsik et al.}{2008b}]{mw2}        Jurcsik J. et al., 2008b, MNRAS, 393, 1553
\bibitem[\protect\citeauthoryear{Jurcsik et al.}{2009}]{kbs1}        Jurcsik J. et al., 2009, MNRAS, 400, 1006
\bibitem[\protect\citeauthoryear{Jurcsik et al.}{2012}]{rzl}         Jurcsik J. et al., 2012, MNRAS, accepted for publication
\bibitem[\protect\citeauthoryear{Koll\'ath}{1990}]{mufran}           Koll\'ath Z., 1990, Occ. Techn. Notes Konkoly Obs., No. 1, http://www.konkoly.hu/staff/kollath/mufran.html
\bibitem[\protect\citeauthoryear{Koll\'ath, Moln\'ar \& Szab\'o}{2011}]{kmsz}           Koll\'ath Z., Moln\'ar, L., Szab\'o, R. 2011, MNRAS, 414, 1111
\bibitem[Kov\'acs(2009)]{ko09} Kov\'acs G., 2009, in Guzik J. A., Bradley P. eds, AIP Conf. Proc. 1170, Stellar Pulsation: Challenges for Theory and Observation, p. 261
\bibitem[\protect\citeauthoryear{Layden}{1994}]{layden} Layden, A., 1994, AJ, 108, 1016
\bibitem[\protect\citeauthoryear{Marconi et al.}{2003}]{marconi}     Marconi M., Caputo F., Di Criscienzo M., Castellani M., 2003, ApJ, 596, 299
\bibitem[Moln\'ar, Koll\'ath \& Szab\'o(2012)]{mol} Moln\'ar L.,  Koll\'ath Z.,  Szab\'o R, 2012, MNRAS,424, 31
\bibitem[\protect\citeauthoryear{Nemec et al.}{2011}]{nemec}         Nemec J.M. et al., 2011, MNRAS, 417, 1022
\bibitem[\protect\citeauthoryear{Papar\'o et al.}{2009}]{paparo}     Papar\'o M., Szab\'o R., Benk\H{o} J. M., Chadid M., Poretti E., Kolenberg K., Guggenberger E., Chapellier E., 2009, in Guzik J. A., Bradley P. eds, AIP Conf. Proc. 1170, Stellar Pulsation: Challenges for Theory and Observation, p. 240
\bibitem[\protect\citeauthoryear{Poretti et al.}{2010}]{poretti}     Poretti E. et al., 2010, A\&A, 520, 108
\bibitem[\protect\citeauthoryear{Schlegel et al.}{1998}]{schlegel}   Schlegel D. J., Finkbeiner D. P., Davis M., 1998, ApJ, 500, 525
\bibitem[\protect\citeauthoryear{Schmidt \& Seth}{1996}]{schmidt}    Schmidt E. G., Seth A., 1996, AJ, 112, 2769
\bibitem[\protect\citeauthoryear{Smolec et al.}{2011}]{sm} Smolec R., Moskalik P., Kolenberg K., Bryson S., Cote M. T., Morris R. L., 2011, MNRAS, 414, 2950
\bibitem[\protect\citeauthoryear{S\'odor}{2012}]{nlfit}              S\'odor \'A., 2012, Occ. Techn. Notes Konkoly Obs., No. 15, http://konkoly.hu/staff/sodor/lcfit.html
\bibitem[\protect\citeauthoryear{S\'odor, Jurcsik \& Szeidl}{2009}]{ip} S\'odor \'A., Jurcsik J., Szeidl B., 2009, MNRAS, 394, 261
\bibitem[\protect\citeauthoryear{S\'odor et al.}{2012}]{aspc}        S\'odor \'A. et al., 2012, to appear in ASPC 
\bibitem[\protect\citeauthoryear{Stothers}{2006}]{stothers}          Stothers R. B., 2006, ApJ, 652, 643
\bibitem[\protect\citeauthoryear{Stothers}{2011}]{stothers2011}      Stothers R. B., 2011, PASP, 123, 127
\bibitem[\protect\citeauthoryear{Szeidl \& Jurcsik}{2009}]{coast}    Szeidl B., Jurcsik J., 2009, Commun. Asteroseis., 160, 17

\end{thebibliography}
\end{document}